\newcommand\apjcls{1}
\newcommand\aastexcls{2}
\newcommand\othercls{3}

\newcommand\papercls{\aastexcls}
\documentclass[tighten, times, twocolumn, trackchanges]{aastex62}
\if\papercls \apjcls
\usepackage{apjfonts}
\else\if\papercls \othercls
\usepackage{epsfig}
\fi\fi
\usepackage{ifthen}
\usepackage{natbib}
\usepackage{amssymb, amsmath}
\usepackage{appendix}
\usepackage{etoolbox}
\usepackage[T1]{fontenc}
\usepackage{paralist}

\if\papercls \apjcls
\newcommand\aas{\ref@jnl{AAS Meeting Abstracts}}% *** added by jh
\newcommand\dps{\ref@jnl{AAS/DPS Meeting Abstracts}}% *** added by jh
\newcommand\maps{\ref@jnl{MAPS}}% *** added by jh
\else\if\papercls \othercls
\usepackage{astjnlabbrev-jh}
\fi\fi

\if\papercls \aastexcls
\hypersetup{citecolor=blue, % color for \cite{...} links
            linkcolor=blue, % color for \ref{...} links
            menucolor=blue, % color for Acrobat menu buttons
            urlcolor=blue}  % color for \url{...} links
\else
\usepackage[%pdftex,      % hyper-references for pdflatex
bookmarks=true,           %%% generate bookmarks ...
bookmarksnumbered=true,   %%% ... with numbers
colorlinks=true,          % links are colored
citecolor=blue,           % color for \cite{...} links
linkcolor=blue,           % color for \ref{...} links
menucolor=blue,           % color for Acrobat menu buttons
urlcolor=blue,            % color for \url{...} links
linkbordercolor={0 0 1},  %%% blue frames around links
pdfborder={0 0 1},
frenchlinks=true]{hyperref}
\fi

\if\papercls \othercls

\else

\fi

\providecommand{\adsurl}[1]{\href{#1}{ADS}}

\makeatletter
\patchcmd{\NAT@citex}
  {\@citea\NAT@hyper@{%
     \NAT@nmfmt{\NAT@nm}%
     \hyper@natlinkbreak{\NAT@aysep\NAT@spacechar}{\@citeb\@extra@b@citeb}%
     \NAT@date}}
  {\@citea\NAT@nmfmt{\NAT@nm}%
   \NAT@aysep\NAT@spacechar\NAT@hyper@{\NAT@date}}{}{}

\patchcmd{\NAT@citex}
  {\@citea\NAT@hyper@{%
     \NAT@nmfmt{\NAT@nm}%
     \hyper@natlinkbreak{\NAT@spacechar\NAT@@open\if*#1*\else#1\NAT@spacechar\fi}%
       {\@citeb\@extra@b@citeb}%
     \NAT@date}}
  {\@citea\NAT@nmfmt{\NAT@nm}%
   \NAT@spacechar\NAT@@open\if*#1*\else#1\NAT@spacechar\fi\NAT@hyper@{\NAT@date}}
  {}{}
\makeatother

\makeatletter
\DeclareRobustCommand{\lowcase}[1]{\@lowcase#1\@nil}
\def\@lowcase#1\@nil{\if\relax#1\relax\else\MakeLowercase{#1}\fi}
\makeatother

\DeclareSymbolFont{UPM}{U}{eur}{m}{n}
\DeclareMathSymbol{\umu}{0}{UPM}{"16}
\let\oldumu=\umu
\renewcommand\umu{\ifmmode\oldumu\else\math{\oldumu}\fi}
\newcommand\micro{\umu}
\if\papercls \othercls
\newcommand\micron{\micro m}
\else
\renewcommand\micron{\micro m}
\fi
\newcommand\microns{\micron}

\let\oldsim=\sim
\renewcommand\sim{\ifmmode\oldsim\else\math{\oldsim}\fi}
\let\oldpm=\pm
\renewcommand\pm{\ifmmode\oldpm\else\math{\oldpm}\fi}
\newcommand\by{\ifmmode\times\else\math{\times}\fi}
\newcommand\ttt[1]{10\sp{#1}}
\newcommand\tttt[1]{\by\ttt{#1}}

\newbox{\wdbox}
\renewcommand\c{\setbox\wdbox=\hbox{,}\hspace{\wd\wdbox}}
\renewcommand\i{\setbox\wdbox=\hbox{i}\hspace{\wd\wdbox}}

\newcount\timect
\newcount\hourct
\newcount\minct
\newcommand\now{\timect=\time \divide\timect by 60
         \hourct=\timect \multiply\hourct by 60
         \minct=\time \advance\minct by -\hourct
         \number\timect:\ifnum \minct < 10 0\fi\number\minct}

\catcode`@=11

\newcommand\comment[1]{}

\newcommand\commenton{\catcode`\%=14}

\renewcommand\math[1]{$#1$}
\newcommand\mathshifton{\catcode`\$=3}

\let\atab=&
\newcommand\atabon{\catcode`\&=4}

\let\oldmsp=\sp
\let\oldmsb=\sb
\def\sp#1{\ifmmode
           \oldmsp{#1}%
         \else\strut\raise.85ex\hbox{\scriptsize #1}\fi}
\def\sb#1{\ifmmode
           \oldmsb{#1}%
         \else\strut\raise-.54ex\hbox{\scriptsize #1}\fi}
\newbox\@sp
\newbox\@sb
\def\sbp#1#2{\ifmmode%
           \oldmsb{#1}\oldmsp{#2}%
         \else
           \setbox\@sb=\hbox{\sb{#1}}%
           \setbox\@sp=\hbox{\sp{#2}}%
           \rlap{\copy\@sb}\copy\@sp
           \ifdim \wd\@sb >\wd\@sp
             \hskip -\wd\@sp \hskip \wd\@sb
           \fi
        \fi}
\def\msp#1{\ifmmode
           \oldmsp{#1}
         \else \math{\oldmsp{#1}}\fi}
\def\msb#1{\ifmmode
           \oldmsb{#1}
         \else \math{\oldmsb{#1}}\fi}

\def\supon{\catcode`\^=7}

\def\subon{\catcode`\_=8}

\def\supsubon{\supon \subon}

\newcommand\actcharon{\catcode`\~=13}

\newcommand\paramon{\catcode`\#=6}

\comment{And now to turn us totally on and off...}

\newcommand\reservedcharson{ \commenton  \mathshifton  \atabon  \supsubon 
                             \actcharon  \paramon}

\catcode`@=12
\reservedcharson

\if\papercls \apjcls

\else

\fi

\newcommand\tnm[1]{\tablenotemark{#1}}
\newcommand\SST{{\em Spitzer Space Telescope}}
\newcommand\Spitzer{{\em Spitzer}}
\newcommand\HST{{\em HST}}
\newcommand\Hubble{{\em Hubble}}
\newcommand\HubbleST{{\em Hubble Space Telescope}}
\newcommand\Webb{{\em James Webb Space Telescope}}
\newcommand\JWST{{\em JWST}}

\newcommand\chisq{\ifmmode{\chi\sp{2}}\else\math{\chi\sp{2}}\fi}
\newcommand\redchisq{\ifmmode{ \chi\sp{2}\sb{\rm red}}
                    \else\math{\chi\sp{2}\sb{\rm red}}\fi}
\newcommand\Teq{\ifmmode{T\sb{\rm eq}}\else$T$\sb{eq}\fi}
\newcommand\Tb{\ifmmode{T\sb{\rm b}}\else$T$\sb{b}\fi}
\newcommand\mjup{\ifmmode{M\sb{\rm Jup}}\else$M$\sb{Jup}\fi}
\newcommand\rjup{\ifmmode{R\sb{\rm Jup}}\else$R$\sb{Jup}\fi}
\newcommand\msun{\ifmmode{M\sb{\odot}}\else$M\sb{\odot}$\fi}
\newcommand\rsun{\ifmmode{R\sb{\odot}}\else$R\sb{\odot}$\fi}
\newcommand\mearth{\ifmmode{M\sb{\oplus}}\else$M\sb{\oplus}$\fi}
\newcommand\rearth{\ifmmode{R\sb{\oplus}}\else$R\sb{\oplus}$\fi}
\newcommand\molhyd{H$\sb{2}$}
\newcommand\methane{CH$\sb{4}$}
\newcommand\water{H$\sb{2}$O}
\newcommand\carbdiox{CO$\sb{2}$}

\newcommand\tnt{$\sp{-2}$}
\newcommand\fraine{F14}
\defcitealias{FraineEtal2014natHATP11bH2O}{\fraine}

\newcommand\vs{\emph{vs.}}
\newcommand\der{\ifmmode{\rm d}\else\math{\rm d}\fi}

\NoNewPageAfterKeywords
\def\hyph{-\penalty0\hskip0pt\relax}

\newcommand\transit{\textsc{Transit}}
\newcommand\BART{\textsc{BART}}
\newcommand\TEA{\textsc{TEA}}
\newcommand\mcc{\textsc{MC3}}
\newcommand\repack{\textsc{repack}}

\shorttitle{BART II: Radiative-transfer Module and HAT-P-11\lowcase{b}}
\shortauthors{Cubillos {\em et al.}}

\begin{document}

\title{
An Open-source Bayesian Atmospheric Radiative Transfer ({\BART}) Code:
II. The {\transit} Radiative-Transfer Module and Retrieval of HAT-P-11b}

\author[0000-0002-1347-2600]{Patricio~E.~Cubillos}
\affiliation{Space Research Institute, Austrian Academy of Sciences,
  Schmiedlstrasse 6, A-8042 Graz, Austria}
\affiliation{Planetary Sciences Group, Department of Physics,
  University of Central Florida, Orlando, FL 32816-2385, USA}

\author[0000-0002-8955-8531]{Joseph~Harrington}
\affiliation{Planetary Sciences Group, Department of Physics,
   University of Central Florida, Orlando, FL 32816-2385, USA}

\author[0000-0002-0769-9614]{Jasmina~Blecic}
\affiliation{Department of Physics, New York University Abu Dhabi,
   PO Box 129188 Abu Dhabi, UAE.}
\affiliation{Planetary Sciences Group, Department of Physics,
   University of Central Florida, Orlando, FL 32816-2385, USA}

\author[0000-0002-9338-8600]{Michael~D.~Himes}
\affiliation{Planetary Sciences Group, Department of Physics,
   University of Central Florida, Orlando, FL 32816-2385, USA}

\author[0000-0002-1607-6443]{Patricio~M.~Rojo}
\affiliation{Departamento de Astronomia, Universidad de Chile,
   Camino del Observatorio, 1515 Las Condes, Santiago, Chile}

\author[0000-0002-4692-4607]{Thomas~J.~Loredo}
\affiliation{Cornell Center for Astrophysics and Planetary Sciences,
   Space Sciences Building, Cornell University, Ithaca, NY 14853-6801, USA}

\author{Nate~B.~Lust}
\affiliation{Planetary Sciences Group, Department of Physics,
   University of Central Florida, Orlando, FL 32816-2385, USA}
\affiliation{Department of Astrophysical Sciences,
   Princeton University, Princeton, NJ 08544, USA}

\author[0000-0002-8211-6538]{Ryan~C.~Challener}
\affiliation{Planetary Sciences Group, Department of Physics,
   University of Central Florida, Orlando, FL 32816-2385, USA}

\author{Austin~J.~Foster}
\affiliation{Planetary Sciences Group, Department of Physics,
   University of Central Florida, Orlando, FL 32816-2385, USA}

\author{Madison~M.~Stemm}
\affiliation{Planetary Sciences Group, Department of Physics,
   University of Central Florida, Orlando, FL 32816-2385, USA}

\author[0000-0003-3077-2127]{Andrew~S.~D.~Foster}
\affiliation{Planetary Sciences Group, Department of Physics,
   University of Central Florida, Orlando, FL 32816-2385, USA}
\affiliation{Cornell Center for Astrophysics and Planetary Sciences,
   Space Sciences Building, Cornell University, Ithaca, NY 14853-6801, USA}

\author[0000-0002-3173-1637]{Sarah~D.~Blumenthal}
\affiliation{Planetary Sciences Group, Department of Physics,
   University of Central Florida, Orlando, FL 32816-2385, USA}
\affiliation{Department of Physics, University of Oxford,
   Oxford OX1 3PU, UK}

\correspondingauthor{Patricio~Cubillos}
\email{patricio.cubillos@oeaw.ac.at}

\comment{
}

\begin{abstract}
This and companion papers by Harrington et al.\ and Blecic et al.\
present the Bayesian Atmospheric Radiative Transfer ({\BART}) code, an
open-source, open-development package to characterize
extrasolar-planet atmospheres.  {\BART} combines a thermochemical
equilibrium abundances ({\TEA}), a radiative-transfer ({\transit}),
and a Bayesian statistical ({\mcc}) module to constrain atmospheric
temperatures and molecular abundances for given spectroscopic
observations.  Here, we describe the {\transit} radiative-transfer
package, an efficient line-by-line radiative-transfer C code for
one-dimensional atmospheres, developed by P. Rojo and further modified
by the UCF exoplanet group.  This code produces transmission and
hemisphere-integrated emission spectra.  {\transit} handles
line-by-line opacities from HITRAN, Partridge \& Schwenke ({\water}),
Schwenke (TiO), and Plez (VO); and collision-induced absorption from
Borysow, HITRAN, and ExoMol.  {\transit} emission-spectra models agree
with models from C. Morley (priv.\ comm.) within a few percent.  We
applied {\BART} to the {\Spitzer} and {\Hubble} transit observations
of the Neptune-sized planet HAT-P-11b.
Our analysis of the combined {\HST} and {\Spitzer} data generally
agrees with those from previous studies, finding atmospheric models
with enhanced metallicity ($\gtrsim 100\times$ solar) and high-altitude
clouds ($\lesssim 1$ mbar level). When analyzing only the {\HST} data,
our models favor high-metallicity atmospheres, in contrast with
the previous analysis by \citeauthor{ChachanEtal2019ajHATP11bPancet} We
suspect that this discrepancy arises from the different choice of
chemistry modeling (free constant-with-altitude {\it vs.}
thermochemical equilibrium) and the enhanced parameter correlations
found when neglecting the {\Spitzer} observations.
%%%
The {\BART} source code and documentation are available
at \href{https://github.com/exosports/BART}
{https://github.com/exosports/BART}.
\end{abstract}


\section{Introduction}
\label{introduction}

Transiting exoplanets are some of the most valuable targets for the
study of planetary atmospheres, since they provide a much more diverse
range of physical properties than that found in the solar-system
planets.  Photometric time-series observations when a planet passes in
front (transits) of its host star constrain the size of the exoplanet;
whereas observations when a planet passes behind (eclipse) constrain
the light emitted and reflected by the planet.  If the planetary mass
is also known, these observations allow us to estimate the bulk
density, gravity, and temperature of the planet's atmosphere.
Furthermore, spectroscopic transit or eclipse observations trace the
transmission or emission spectral variations imprinted by the
atmospheric species, respectively.  Since each species in a planet's
atmosphere produces a specific absorption and emission pattern (at a
given temperature and pressure), with enough spectral resolution and
coverage one can constrain the composition of the exoplanetary
atmosphere.

Numerous physical processes can shape the temperature and composition
of an atmosphere, ranging from formation
scenarios \citep[e.g.,][]{ObergEtal2011apjCOsnowlines,
DrummondEtal2019mnrasCOratioSpectra}; equilibrium-chemistry,
photochemistry,
kinetics \citep[e.g.,][]{MosesEtal2013apjCompositionalDiversity,
MadhusudhanEtal2016ssrChemistryFormation}; circulation
dynamics \citep[e.g.,][]{ShowmanEtal2009apjRadGCM,
MayneEtal2014aaGCMunifiedModel}; to cloud
physics \citep[e.g.,][]{MarleyEtal2013cctpCloudsHazes,
MorleyEtal2015apjFlatSpectra}.  However, once the properties of an
atmosphere are known or assumed (i.e., the pressure, temperature, and
composition), the radiative-transfer equation ultimately relates the
atmospheric state to the observed spectrum.

Radiative-transfer calculations for exoplanet atmospheres need to deal
with a wide range of challenges, including limited laboratory or
theoretical opacity data at high
temperatures \citep{FortneyEtal2016whiteLabDataExoplanets},
oversimplified atmospheric
models \citep[e.g.,][]{RocchettoEtal2016apjJWSTbiases,
CaldasEtal2019aaTransmission3Deffects,
TaylorEtal2020mnrasEmissionBiases}, or stellar
contamination \citep{RackhamEtal2017apjStellarHeterogeneityI}.

Radiative transfer models for substellar-object atmospheres have been
continuously developed for decades, with early exoplanet models
naturally building upon the previous experience from Solar System
planets and brown dwarfs \citep[e.g.,][]{MckayEtal1989icarTitan,
MarleyEtal1996sciAtmosphereGliese229B,
OppenheimerEtal1998apjGliese229B,
BurrowsEtal1997apjEGPandBrownDwarfs}.  Consequently, to date there are
a large number of models, varying in complexity and focus, from
1D \citep[e.g.,][]{BurrowsEtal2005apjSecondaryEclipseModels,
Fortney2005mnrasClouds, SeagerEtal2005apjThermalEmission} to
3D \citep[e.g.,][]{FortneyEtal2010apj3DTransmission}, or focused on
infrared \citep[e.g.,][]{Dudhia2017jqsrtReferenceForwardModel} or
optical
wavelengths \citep[e.g.,][]{LupuEtal2016ajOpticalAtmosphericRetrieval}.

The limited signal-to-noise ratio, spectral resolution, and spectral
coverage of exoplanet data generally leads to poorly constrained and
degenerate constraints of the atmospheric properties.  This has led to
the development of high-performance radiative-transfer
codes \citep[e.g.,][]{WaldmannEtal2015apjTauRexI,
KemptonEtal2017paspExoTransmit, MalikEtal2017ajHELIOS} and
statistically-robust Bayesian retrieval
frameworks \citep[e.g.,][]{MadhusudhanSeager2010apjRetrieval,
BennekeSeager2012apjRetrieval, LineEtal2013apjRetrievalI,
WaldmannEtal2015apjTauRexI, LavieEtal2017HeliosR,
LupuEtal2016ajOpticalAtmosphericRetrieval,
BarstowEtal2017apjHotJupiterRetrieval,
EvansEtal2017natWASP121bEmissionHST, 
MacDonaldMadhusudhan2017mnrasRetrievalHD209b,
GandhiMadhusudhan2018mnrasHyDRA,
MolliereEtal2019aaPetitRADTRANS,
ZhangEtal2019paspPLATON,
LothringerBarman2020ajPETRA,
MinEtal2020aaARCiS,
CubillosBlecic2021mnrasPyratBay}---some of which are available as open-source
software.  A Bayesian retrieval is the most appropriate approach when
fitting poorly-constrained models.  This approach explores the
parameter space by evaluating a large number of models.  Here, the
data drive the exploration towards the most-probable solutions, guided
by the likelihood function and priors.  The data quality determines
the span of the highest-probability-density credible region of the
model parameters (i.e., the error bars) and quantify any correlations.
Weakly constraining data produce large model parameter uncertainties.

Nowadays, exoplanet atmospheric characterization is developing at a
thrilling pace.  The combined observations from ground-based,
{\Spitzer}, and {\Hubble} telescopes create a high demand for
interpretation of the data.  Furthermore, the
{\Webb} \citep[{\JWST},][]{GardnerEtal2006ssrJWST} will continue this
trend, allowing for exoplanet characterization with unprecedented
detail.  The limited 5--10 year mission duration {\JWST}, will require
prompt modeling of the data, hopefully from multiple, independent
teams.  Having a range of retrieval frameworks readily available
benefits the community by facilitating preparatory and
inter-comparison studies to fully exploit the constraining power of
present and future instrumentation.

\subsection{The Bayesian Atmospheric Radiative-Transfer Package}
\label{sec:bart}

In this context, we developed the open-source Bayesian Atmospheric
Radiative Transfer ({\BART}) package for statistically robust
exoplanet characterization, which we present jointly in three
articles.  \citet[][submitted]{HarringtonEtal2021psjBART} presents an
overview of the retrieval framework, testing, and an application to
the eclipse data of HD~189733\,b.  This article focuses on the
radiative-transfer module, with an application to the transit data of
HAT-P-11b.  \citet[][submitted]{BlecicEtal2021psjBART} focuses on the
initialization and post-processing routines, with an application to
the eclipse data of WASP-43b.  {\BART} offers both forward-model and
retrieval tools to model exoplanet spectra and constrain the
atmospheric temperature and composition.  {\BART} is an
open-development project available under the Reproducible Research
Software License \citep[see][]{HarringtonEtal2021psjBART}.  The code
repositories and documentation are available under version control
at \href{https://github.com/exosports/BART}
{{https://github.com/exosports/BART}}.

{\BART} combines three general-purpose and independent sub-packages.
The Thermochemical Equilibrium Abundances
code \citep[{\TEA},][]{BlecicEtal2016apsjTEA} computes mole mixing
ratios (abundances) of atmospheric species under thermochemical
equilibrium.  Thus, {\TEA} can provide an initial guess for the
atmospheric abundances to the {\BART} framework.  The
radiative-transfer package, {\transit}, computes line-by-line
transmission or emission spectra for given one-dimensional atmospheric
models.  The Multi-Core Markov-Chain Monte Carlo statistical
package \citep[{\mcc},][]{CubillosEtal2017apjRednoise} is a
general-purpose, model-fitting package that provides parallel
multiprocessor Markov-chain Monte Carlo (MCMC) to explore the
parameter space. {\mcc} implements the snooker Differential-evolution
MCMC algorithm \citep[DEMC,][]{terBraak2006DifferentialEvolution,
terBraak2008SnookerDEMC}.

This paper is organized as follows: section \ref{sec:rt}
details the radiative-transfer calculation, its assumptions, geometry
treatment, available data bases, and validation tests.  Section
\ref{sec:analysis} shows our retrieval analysis of the extrasolar
planet HAT-P-11b using {\BART}.  Finally,
Section \ref{sec:conclusions} presents our prospects for the {\BART}
project and summarizes our conclusions.

\section{The {\transit} Radiative-transfer Package}
\label{sec:rt}

Here we describe the treatment of the radiative-transfer equation in
the {\transit} code, which is generally based
on \citet{Andrews2000bookAtmosphericPhysicsIntro}
and \citet{RybickiLightman1986bookRadiativeProcesses}.  The
radiative-transfer equation describes how light propagates as it
travels through a medium.  Let the specific intensity, $I\sb{\nu}$,
denote the power carried by rays per unit area, $\der A$, per unit
wavenumber, $\nu$, in the interval $\der \nu$, within a solid angle
$\der\Omega$.  Then, the radiative-transfer equation for the specific
intensity is given by:
\begin{equation}
  \frac{{\rm d}I\sb{\nu}}{{\rm d}s} = -e\sb{\nu}(I\sb{\nu} - S\sb{\nu}),
\label{eq:RadTran}
\end{equation}
where $s$ is the path traveled by the light ray, $e\sb{\nu}$ is the
extinction coefficient, and $S\sb{\nu}$ is the source function.  The
extinction depends on the atmospheric composition, pressure, and
temperature.  The observed flux spectrum is the integral of the
specific intensity multiplied by the cosine of the polar angle over the solid angle.  Therefore, by solving the
radiative-transfer equation for a given observing geometry and
atmospheric model, we can model the observed spectrum as a function of
wavelength.

The {\transit} package solves the one-dimensional radiative-transfer
equation for two relevant cases of exoplanet observations: the
transmission spectrum for transit observations, and the
hemisphere-integrated emission spectrum for eclipse observations.  The
model assumes hydrostatic balance, local thermodynamic equilibrium,
and the ideal gas law.  The opacity comes from electronic, rotational,
and vibrational line-transition absorptions (hereafter, simply called
``line transitions'', LT), collision-induced absorption (CIA), Mie and
Rayleigh scattering, bound-free and free-free transitions (e.g.,
H$^-$), and aerosols.  The {\transit} code currently implements
line-transition, CIA, Rayleigh, and gray-cloud opacities.

{\transit} is a C, modular, object-oriented code, wrapped with
SWIG\footnote{\href{http://swig.org/}{swig.org/}} for use in Python.
The code was originally developed by \citet{Rojo2006phdThesis}.  We
rewrote the code to be one of the core packages of the {\BART}
project, implementing the eclipse-geometry mode, and improving the
overall performance.  The current version of {\transit} is an
open-source project hosted
at \href{https://github.com/exosports/transit}
{https://github.com/exosports/transit}.  The repository includes a
user manual that describes the main features, inputs, and outputs of
the code, and provides sample runs.  A second document aimed at
developers, the code manual, details the data structures and the file
formats.

\subsection{Transit Workflow}
\label{sec:workflow}

\begin{figure}[tb]
\centering
\includegraphics[width=\linewidth, clip]{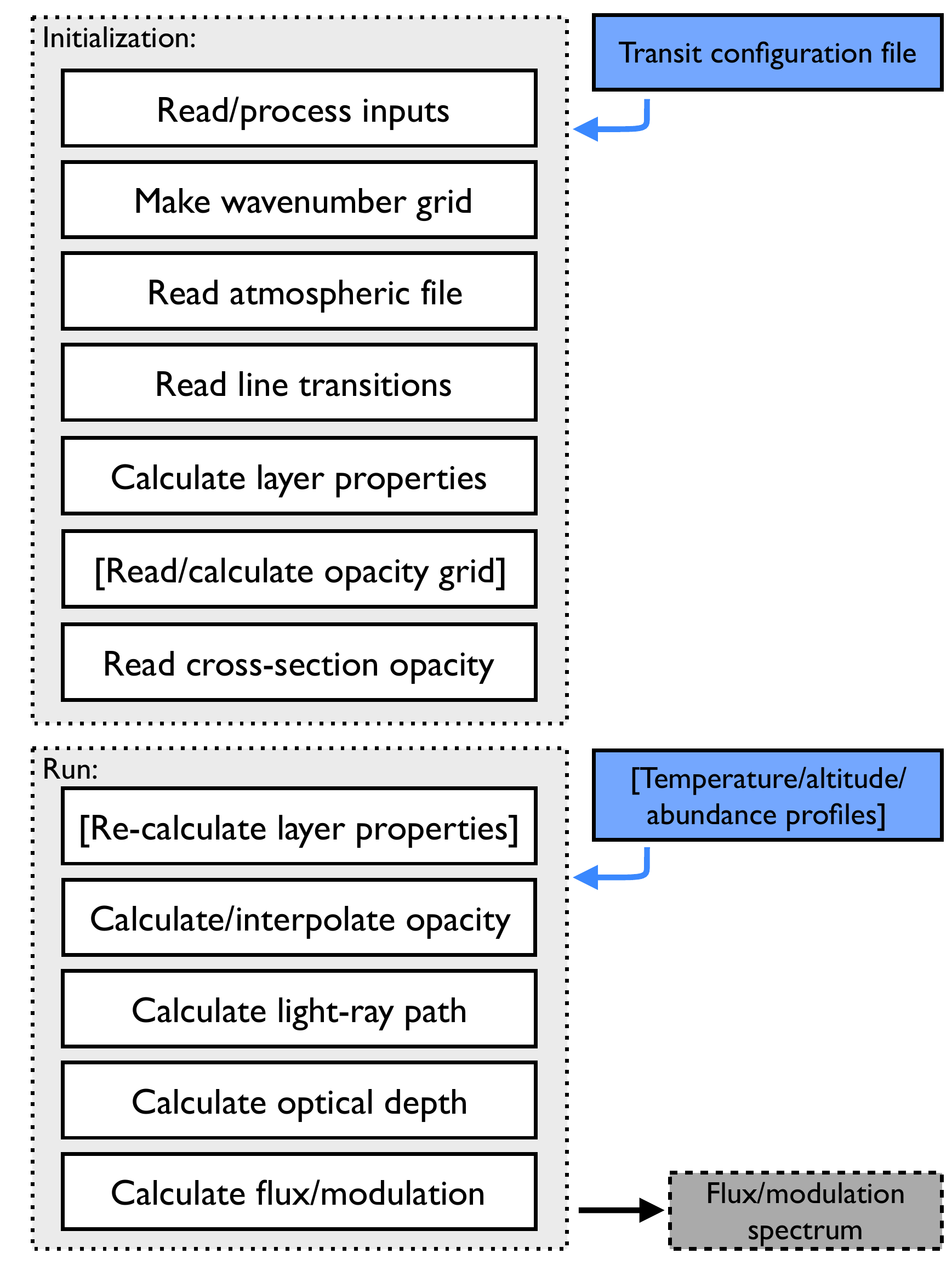}
\caption{{\transit} flow chart.  The white
  boxes denote the main routines, the blue boxes the code inputs, and
  the gray box the output spectrum file.  {\transit} sequentially
  executes the routines from the top down.  The items in brackets are
  optional steps.  The execution consists of two main steps.  The
  initialization step reads the input files, creates the (equi-spaced)
  wavenumber array, calculates the layer's altitude and species
  density and partition function, and computes or reads a tabulated
  opacity grid (optional).  The second main step updates the
  atmospheric model (if requested) and computes the atmospheric
  opacity over the spectral range, the light-ray path, the optical
  depth, and the intensity and flux spectra.  Once the initialization
  is executed, the user can produce multiple spectrum calculations for
  desired atmospheric models, for example, during an MCMC the code
  cycles over the bottom section.}
\label{fig:TransitFlow}
% SOURCE: /home/pcubillos/Dropbox/latex/2016_bart/pc/current/figs/TransitFlow.key
\end{figure}

The {\transit} program divides the spectrum calculation into two main
steps (Fig.\ \ref{fig:TransitFlow}). The first step initializes the
required variables, reading and processing the input files.  The user
can optionally create an opacity grid to avoid computing line-by-line
opacities for each spectrum calculation.  The second step solves the
radiative-transfer equation, which involves computing the ray paths,
calculating the opacity at the given pressure, temperature, and
wavenumber, and producing the spectrum.  This separation is intended
to improve the performance of the code when run multiple times (e.g.,
in a MCMC).  With the exception of the atmospheric profiles, none of
the inputs read during the initialization change during the course of
the calculations.  Thus, the user can produce multiple spectrum
models, updating the atmospheric model each time, without the need to
reprocess the input files (which can be large, like the line
transition information).

{\transit} requires as inputs: (1) a configuration file that indicates
the wavenumber sampling, observing geometry, input files, etc.; (2) a
one-dimensional atmospheric model that specifies the atmospheric
composition and the pressure, altitude, temperature, and species
abundances of each layer; and (3) line-by-line and/or cross-section
opacity files.  The following sections further detail the assumptions
and available opacity databases for {\transit}.

\subsection{Observing Geometry}
\label{sec:geometry}

The {\transit} code computes transmission and emission spectra,
appropriate to exoplanet transit or secondary-eclipse observations,
respectively.  Depending on the observing geometry, the code makes
different approximations about the geometry and fluxes.
Much of the formulation laid out in this section is more thoroughly
described in textbooks \citep[see e.g.,][and references
therein]{Seager2010bookExoplanetAtmospheres}.

\subsubsection{Transit Geometry}
\label{sec:transit}

\begin{figure}[tb]
\centering
\includegraphics[width=\linewidth, clip, trim=0 40 0 40]
{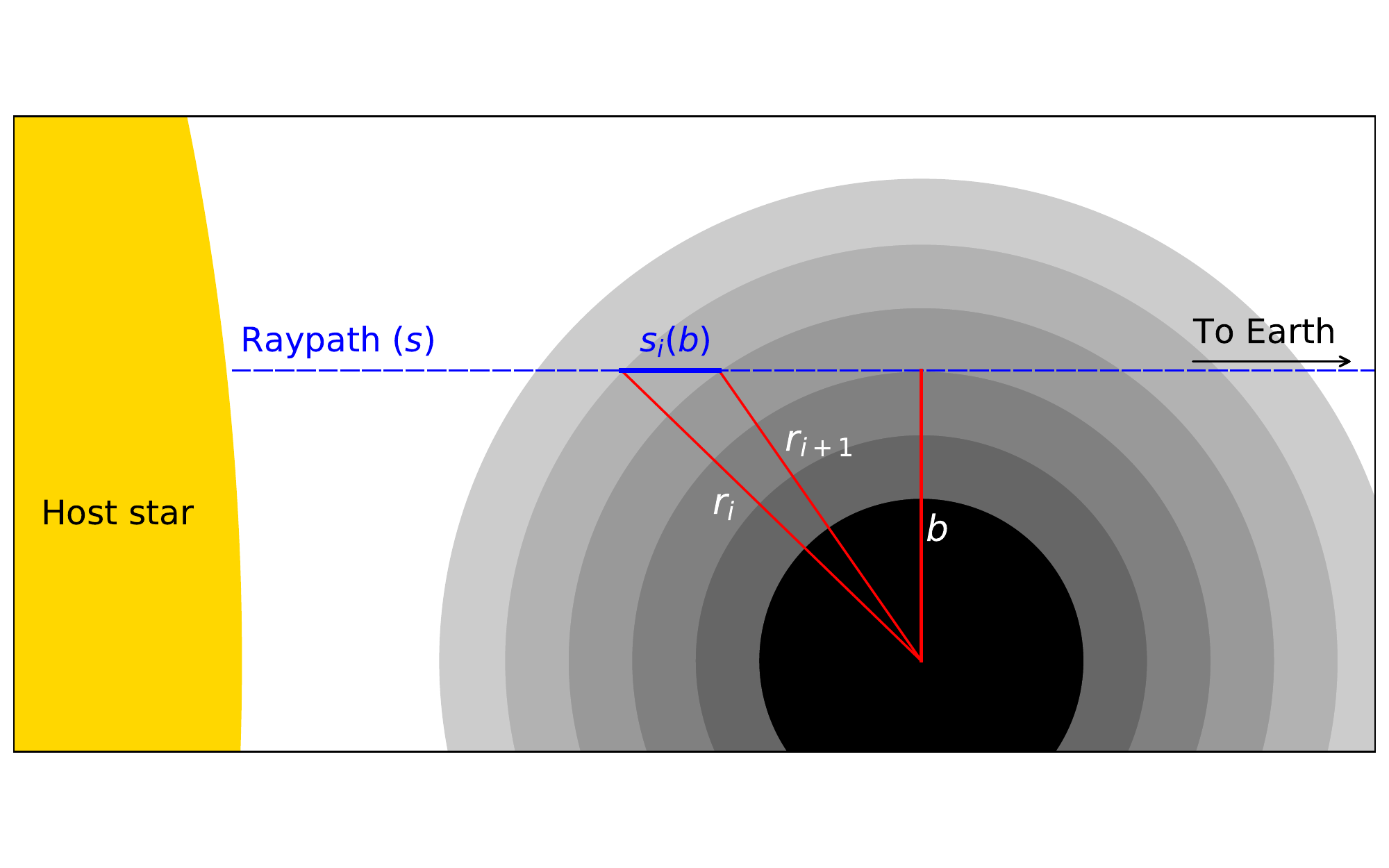}
\caption{
Transit-geometry observation diagram.  The atmospheric model consists
of a set of one-dimensional spherically-symmetrical shells (gray
gradient) around the planet center.  All the observed light rays (blue
dashed line) travel nearly parallel to each other and, hence, the
optical depth for each ray depends exclusively on the impact parameter
($b$), where the path traveled between two layers (blue solid line) is
given by $s_i(b) = \sqrt{r_{i}^2-b^2} - \sqrt{r_{i+1}^2-b^2}$.}
\label{fig:transitSketch}
% SOURCE: /home/pcubillos/ast/compendia/CubillosEtal2021_HATP11b/figure_transit.py
\end{figure}

During a transit event (Fig.\ \ref{fig:transitSketch}) the planet
blocks a fraction of the stellar light, which is proportional to the
planet-to-star area ratio, $R\sb{\rm p}\sp{2}$/$R\sb{\rm s}\sp{2}$.
Since each species imprints a characteristic absorbing pattern as a
function of wavelength, the planetary atmosphere modulates the
transmission (or modulation) spectrum at each wavenumber $\nu$:
\begin{equation}
  M\sb{\nu}  = \frac{F\sb{\nu,O}-F\sb{\nu,T}}{F\sb{\nu,O}},
  \label{eq:modulation}
\end{equation}
where $F\sb{\nu,T}$ and $F\sb{\nu,O}$ are the observed specific fluxes
in and out of transit, respectively.

For transmission geometry, the planetary emission is small compared to
the stellar intensity.  Then, we can neglect the source-function term
from the radiative-transfer equation. Eq.\ (\ref{eq:RadTran}) then
leads to:
\begin{equation}
  \frac{{\rm d}I\sb{\nu}}{{\rm d}s} = -e\sb{\nu}I\sb{\nu}.
\end{equation}

Let $I\sb{\nu}\sp{0}$ be the stellar specific intensity.  By defining
the optical depth along a path, $s$, as:
\begin{equation}
\label{eq:tau}
\tau\sb{\nu} = \int\sb{\rm path} e\sb{\nu} {\der}s,
\end{equation}
the solution for the transmission radiative-transfer equation becomes:
\begin{equation}
I\sb{\nu} = I\sb{\nu}\sp{0} e\sp{-\tau\sb{\nu}}.
\end{equation}

The specific flux in the direction of the observer, $F\sb{\nu}$, is
the integral of the first moment of the specific intensity
over the solid angle, $\der\Omega$:
\begin{equation}
F\sb{\nu} = \int I\sb{\nu} cos\theta \der\Omega,
\label{eq:SpecificFlux}
\end{equation}
where $\theta$ is the angle between the ray beams and the normal
vector of the detector.  Since the distance from Earth to the system,
$d$, is much larger than the distance from the star to the planet, the
stellar radius, and the planetary radius ($d \gg a$, $d \gg R\sb{\rm
s}$, and $d \gg R\sb{\rm p}$, respectively), we can assume that the
observed rays travel in a parallel beam thorough the planetary
atmosphere.  Considering an idealized star with a uniform spectrum
across its surface (i.e., ignoring surface features and accounting for
limb darkening in the analysis producing the transit depths) and a
spherically symmetric planet, we can consider the planet centered in
front of the star and rewrite the solid-angle integral as an integral
over the projected disks of the planet and the star:
\begin{equation}
F\sb{\nu}  = \frac{2\pi}{d\sp{2}} \int\sb{0}\sp{R\sb{\rm s}} I\sb{\nu} r \der r.
\label{eq:tranflux}
\end{equation}

For the out-of-transit flux, Equation (\ref{eq:tranflux}) gives:
\begin{equation}
F\sb{\nu,O}  = \pi I\sb{\nu}\sp{0} \left(\frac{R\sb{\rm s}}{d}\right)\sp{2}.
\label{eq:ootFlux}
\end{equation}

For the in-transit flux, we assume that the planetary atmosphere is a
set of spherically-symmetrical homogeneous layers
(Fig.\ \ref{fig:transitSketch}).  Then, the optical depth becomes a
function of the impact parameter of the light ray, $b$: $\tau\sb{\nu}
= \tau\sb{\nu}(b)$.  Considering now a planetary altitude, $R\sb{\rm
top}$, high enough such that $e\sp{-\tau\sb{\nu}(R\sb{\rm
top})} \approx 1$ (in practice, the top layer of the atmospheric
model), we can separate the integral of Eq.\ (\ref{eq:tranflux}) into
the regions occulted and unocculted by the planet.  Further,
considering a limb-darkening-corrected transit depth, the in-transit
specific flux becomes:
\begin{eqnarray}
F\sb{\nu,T} & = & \frac{2\pi}{d\sp{2}} 
   \int\sb{0}\sp{R\sb{\rm s}} I\sb{\nu}\sp{0} e\sp{-\tau\sb{\nu}(b)}r{\der}r, \\
            & = & \frac{2\pi}{d\sp{2}} I\sb{\nu}\sp{0}
   \left[ \int\sb{0}\sp{R\sb{\rm top}} e\sp{-\tau\sb{\nu}} b \der b\ +\
        \frac{R\sb{\rm s}\sp{2} - R\sb{\rm top}\sp{2}}{2} \right].
\label{eq:itFlux}
\end{eqnarray}

Finally, substituting Equations (\ref{eq:ootFlux}) and
(\ref{eq:itFlux}) into (\ref{eq:modulation}), we calculate the
modulation spectrum as:
\begin{equation}
M\sb{\nu}  = \frac{1}{R\sb{\rm s}\sp{2}}\left[ R\sb{\rm top}\sp{2} - 
                2\int\sb{0}\sp{R\sb{\rm top}} e\sp{-\tau\sb{\nu}} b \der b
             \right].
\end{equation}

\subsubsection{Eclipse Geometry}
\label{sec:eclipse}

An eclipse event reveals the planetary emission integrated over the
day-side hemisphere.  In this case, {\transit} obtains the emerging
intensity at the top of the atmospheric model by solving the
radiative-transfer equation under the plane-parallel approximation.
Additionally, {\transit} adopts the local thermodynamic equilibrium
approximation, where the source function becomes the Planck function
$B\sb{\nu}(T)$.

Considering the vertical optical depth, $\der \tau\sb{\nu,z} =
-e\sb{\nu} \der z$ (with origin $\tau=0$ at the top of the
atmosphere), the path, $\der s$, of a ray with an angle $\alpha$, with
respect to the normal vector, is related to the vertical path as $\der
s = \der z/\cos\alpha \equiv \der z / \mu$.  Then, the
radiative-transfer equation becomes:
\begin{equation}
-\mu\frac{\der I\sb{\nu}}{\der \tau\sb{\nu,z}} = - I\sb{\nu}  +  B\sb{\nu},
\end{equation}
which can be rewritten as:
\begin{equation}
-\mu\frac{\der}{\der \tau\sb{\nu,z}}\left( I\sb{\nu}e\sp{-\tau\sb{\nu,z}/\mu} \right) 
 =  B\sb{\nu}e\sp{-\tau\sb{\nu,z}/\mu}.
\label{eq:RTeclipse}
\end{equation}

{\transit} calculates the emergent intensity by integrating Equation
(\ref{eq:RTeclipse}) from the deep layers to the top of the
atmosphere.  At depth, the atmosphere is optically thick
($\tau=\tau\sb{b} \gg 1$), such that $\exp(-\tau\sb{b}/\mu) \to 0$.
Therefore, the emergent intensity at the top of the atmosphere is
given by:
\begin{equation}
I\sb{\nu}(\tau=0, \mu) = 
     \int\sb{0}\sp{\tau\sb{b}}B\sb{\nu}e\sp{-\tau/\mu}\der\tau/\mu.
\label{eq:EclipseIntensity}
\end{equation}

By changing variables from the angle $\theta$ to the angle on the
planet hemisphere, $\alpha$ (related by $d \sin\theta =
R\sb{\rm p}\sin\alpha$),  the
emergent specific flux, Eq.\ (\ref{eq:SpecificFlux}), becomes:
\begin{eqnarray}
F\sb{\nu} & = & \left(\frac{R\sb{\rm p}}{d}\right)\sp{2} 
                \int\sb{0}\sp{2\pi}\int\sb{0}\sp{\pi/2} 
                   I\sb{\nu} \cos(\alpha)\sin(\alpha) \der\alpha \der\phi \\
          & = & 2\pi \left(\frac{R\sb{\rm p}}{d}\right)\sp{2}
                  \int\sb{0}\sp{1} I\sb{\nu} \mu\der\mu.
\end{eqnarray}

{\transit} approximates this integral by summing over a discrete set
of angles (user input), sampling from the sub-stellar point to the
terminator:
\begin{equation}
F\sb{\nu} \approx \pi \left(\frac{R\sb{\rm p}}{d}\right)\sp{2}
           \sum\sb{i} I\sb{\nu}(\mu\sb{i}) \Delta\mu\sb{i}\sp{2},
\label{eq:pflux}
\end{equation}
where the boundaries for the spans $\Delta\mu\sb{i}\sp{2}$ are
calculated from the mean points between consecutive angles.  By
default the code adopts five angles at 0$^\circ$, 20$^\circ$,
40$^\circ$, 60$^\circ$, and 80$^\circ$.  We found no significant
differences when varying the values or when comparing with a Gaussian
quadrature procedure of a similar number of points.  Finally,
{\transit} computes $I\sb{\nu}(\mu\sb{i})$ through a Simpson numerical
integration of Equation (\ref{eq:EclipseIntensity}) and returns the
emergent specific flux as measured on the surface of the planet
($d \equiv R\sb{\rm p}$) in units of erg\,s$\sp{-1}$cm$\sp{-2}$cm:
\begin{equation}
  F\sb{\nu}\sp{\rm p} \approx
     \pi \sum\sb{i} I\sb{\nu}(\mu\sb{i}) \Delta\mu\sb{i}\sp{2}.
\end{equation}

\subsection{Atmospheric Extinction}

There are multiple sources of atmospheric opacity, each one with their
particular properties.  The current code incorporates the main sources
of opacity considered for exoplanet atmospheres: collision-induced
absorption, line-transition opacities, Rayleigh scattering, and
clouds.

\subsubsection{Collision-induced Absorption}

Collision-induced absorption is one of the main sources of atmospheric
opacity.  CIA occurs when particles without an intrinsic electric
dipole moment collide.  The collisions induce a transient dipole
moment, which allows dipole transitions.  The short interaction time
of the collisions broadens the line profiles, generating a smooth CIA
spectrum.  The CIA opacity scales with the density of the colliding
species, and thus becomes more relevant at the deeper, higher-pressure
layers of the atmosphere \citep{SharpBurrows2007apjOpacities}.  For
gas-giant planets, the two most important CIA sources are
{\molhyd}--{\molhyd} and, to a lesser extent, {\molhyd}--He
collisions.  For secondary-atmosphere planets, the dominant CIA source
will depend on the most abundant species in the given
atmosphere.  \citet{KarmanEtal2019icarHITRANupdateCIA} provides CIA
data of several molecular pairs of interest for primary and secondary
atmospheres.

\subsubsection{Line-transition Absorption}

Line transitions arise when a species absorbs or emits photons at
specific wavelengths, corresponding to the characteristic energies
between its electronic, rotational, and vibrational quantum levels.
The atmospheric temperature and pressure affect the strength and shape
of the line transitions.

{\transit} calculates the line-transition extinction coefficient,
$e\sb{\nu}$, in a line-by-line scheme, adding the contribution from
each broadened line-transition, $j$, as:
\begin{equation}
e\sb{\nu} = \sum\sb{j} S\sb{j} V(\nu-\nu\sb{j}),
\label{eq:opacity}
\end{equation}
where $\nu\sb{j}$ is the wavenumber of the line transition, $S\sb{j}$
is the line strength (in cm{\tnt}), and $V$ is the line profile
(Voigt).  The line strength is given by:
\begin{equation}
\small
S\sb{j} = \frac{\pi e\sp{2}}{m\sb{e}c\sp{2}} \frac{(gf)\sb{j}}{Z\sb{i}(T)}
          n\sb{i} \exp\left(-\frac{h c E\sb{\rm low}\sp{j}}{k\sb{B}T}\right)
            \left\{1-\exp\left(-\frac{hc\nu\sb{j}}{k\sb{B}T}\right)\right\},
\end{equation}
\normalsize
where $gf\sb{j}$ (unitless) and $E\sp{j}\sb{\rm low}$ (in cm$^{-1}$)
are the weighted oscillator strength and lower-state energy level of
the line transition, respectively; $Z\sb{i}$ and $n\sb{i}$ are the
partition function and number density of the isotope $i$,
respectively; $T$ is the atmospheric temperature; $e$ and $m\sb{e}$
are the electron's charge and mass, respectively; $c$ is the speed of
light, $h$ is Planck's constant; and $k\sb{\rm B}$ is the Boltzmann's
constant.

The Voigt profile considers the Doppler and the Michelson-Lorentz
collision broadening \citep[though, neglecting J{\hyph}dependence,
e.g.,][]{BartonEtal2017jqsrtH2Obroadening}.  The Doppler and Lorentz
half-widths at half
maximum \citep[HWHM,][]{Goody1996AtmosphericPhysics} are,
respectively:
\begin{eqnarray}
\label{eq:alphad}
\sigma & = &
    \frac{\nu\sb{j}}{c} \sqrt{\frac{2k\sb{\rm B}T\ln{2}}{m\sb{i}}}, \\
\label{eq:alphal}
\gamma & = &
    \frac{1}{c} \sum\sb{a} n\sb{a}d\sb{a}\sp{2}
        \sqrt{\frac{2k\sb{\rm B}T}{\pi}
        \left(\frac{1}{m\sb{i}} + \frac{1}{m\sb{a}}\right)},
\end{eqnarray}
where the sub-index $i$ refers to the absorbing species, the sub-index
$a$ of the sum runs over all species in the atmosphere.  $m_i$ and
$m_a$ are the masses of the species $i$ and $a$ (respectively),
$d_a=r_i+r_a$ is the collision diameter between the interacting
particles (with $r_i$ and $r_a$ the collision radii of species $i$ and
$a$, respectively), and $n_a$ the number density of species $a$.
{\transit} computes the Voigt profiles following the algorithm
of \citet[][]{Pierluissi1977jqsrtVoigt}.
%%%
The wings of these line profiles are truncated at a user-defined
cutoff value proportional to the Voigt HWHM of the line (typically
100--500 HWHM). We note that this is an {\it ad hoc} choice given the
general lack of knowledge on this topic \citep[e.g., see the
discussion by][]{GharibNezhadEtal2021apjsExoplines}.

Depending on the wavelength range and input opacity databases, a
radiative-transfer calculation typically involves from thousands to
billions of line transitions.  Thus, the line-by-line opacity
calculation is the most computationally intensive task of the process,
in particular computing the Voigt profiles.  Noting that the typical
width of the Lorentz HWHM, being inversely proportional to the species
number density, hence to the pressure, increases several orders of
magnitude from the top to the bottom of the atmosphere.  The Voigt
profile is then dominated by the Doppler profile at the top of the
atmosphere, and by the Lorentz profile at the bottom of the
atmosphere.  To improve performance, we pre-compute a 3D table of
Voigt profiles \citep[see also Eqs.\ (2)--(11)
of][]{HarringtonEtal2021psjBART} with axes of wavenumber, Doppler
width, and Lorentz width, covering the width ranges possible due to
the pressures, temperatures, and species in the calculation.  The
sampling ranges and rates are user parameters, with defaults far
better than needed for accuracy in current exoplanet retrievals (i.e.,
thousands of samples per wavenumber).  Creating a high-resolution
lookup table once is much quicker and more accurate than either
calculating the profile per line or interpolating a lower-resolution
table for each line, and, once created, accessing it takes the same
time regardless of size, within reason.

The output opacity array has coarser sampling, typically 1 cm$^{-1}$
for current exoplanet retrievals, although this value can be set by
the user.  When filling this array, for each line, {\transit} selects
the Voigt profile with the closest width parameters and shifts it by
an integer number of profile fine samples to best match the line peak.
It then samples values from the profile at the output grid's
wavenumbers.

Although, given to the nonlinearity of the radiative-transfer
calculation, simply sampling from the finer Voigt profiles may bias
the opacity contribution of individual lines, due to the combined
contributions of the millions of lines the overall bias averages out.
{\citet{Rocchetto2017phdThesis}} showed that for current low- and
mid-resolution observations (e.g., {\Spitzer} or {\HST}), this
approximation introduces a negligible opacity bias on model spectra at
output sampling rates on the order of 1 cm$^{-1}$.  We confirmed this
during the implementation of the {\transit} code, and also found that
sampling opacity produces more accurate spectra than averaging opacity
over bins.  Certainly, these biases will be more severe at longer
wavelengths, e.g., for {\JWST}\//MIRI observations.  Thus, modeling
such observations will require a finer sampling.

Additionally, {\transit} incorporates a line-strength cutoff (an
adjustable parameter) that prevents the computation of the weaker
lines that do not significantly contribute to the opacity.  Given the
large span of the plausible temperatures and pressures in the
atmosphere, the relative strength of different lines can change
significantly from layer to layer.  Therefore, {\transit}'s cutoff
threshold is relative to the largest line strength in each layer,
rather than a fixed cutoff.  To further minimize the number of
evaluated Voigt profiles, the code automatically adds the line
strengths of transitions from the same isotope that fall in a same
wavenumber bin.

Lastly, {\transit} provides the option to pre-calculate a grid of the
line-by-line opacities during the initialization step.  The opacity
grid is a four-dimensional table that contains the opacities (in
cm$\sp{2}$ g$\sp{-1}$) evaluated over the wavenumber array, at each
atmospheric pressure level, for a grid of temperatures, and for each
absorbing species.  The opacity grid speeds up the spectrum evaluation
allowing {\transit} to interpolate the opacities from the table,
instead of repeatedly computing the line-by-line calculations.
Including the opacity grid is necessary to run an atmospheric
retrieval in a reasonable amount of time (hours to days), since an
MCMC typically requires on the order of millions of spectrum
evaluations.  These grids are computed at the wavelengths and
pressures desired for a given retrieval, and thus the procedure only
needs to interpolate in temperature (linear interpolation).  Note that
the execution time varies widely depending on the wavelength range,
spectral sampling rate, number of atmospheric layers, number of
spectroscopically-active species, and of course, the hardware.

\subsubsection{Clouds and Rayleigh Scattering}
\label{sec:clouds}

The {\transit} code implements Rayleigh scattering via the parametric
model of \citet{LecavelierEtal2008aaRayleighHD189733b}, where the
scattering cross section is given by $\sigma_{\rm ray}
= \sigma_0(\lambda/\lambda_0)^{-4}$, with $\sigma_0$ and $\lambda_0$
adjustable parameters.  Under ideal gas law, the Rayleigh extinction
coefficient is given by:
\begin{equation}
e_{\rm ray} = \kappa_{\rm ray} e_0 \frac{p}{T} \lambda^{-4},
\end{equation}
where $\kappa_{\rm ray}$ is a fitting parameter, and the constant
$e_0=4.91$\tttt{$-23$} K\,bar$^{-1}$\,cm$^{3}$ is chosen such that
$e_{\rm ray}$ matches the {\molhyd} Rayleigh scattering of a
solar-composition atmosphere when $\kappa_{\rm ray}=1$.

The code also implements the gray cloud-deck model that has been
largely used by most atmospheric retrievals.  This model is
parameterized by a cloud-top pressure ($p_{\rm cloud}$) below which
the atmosphere becomes instantly optically thick.

\subsection{Opacity Databases}
\label{sec:databases}

The {\transit} package handles two types of input opacity databases,
cross-section (CS, opacity as a function of wavenumber and
temperature) and line-by-line data (LBL, sets of individual line
transition opacity parameters).  Table \ref{table:opacity} list the
opacity databases currently available for {\transit}.

{\renewcommand{\arraystretch}{1.0}
\begin{table}[ht]
\centering
\caption{{\transit} opacity databases}
\label{table:opacity}
\footnotesize
\begin{tabular*}{\linewidth} {@{\extracolsep{\fill}} lccc}
\hline
\hline
Source     & Species & Type & Format \\
\hline
HITRAN/HITEMP\tnm{a} & {\water}, CO, {\carbdiox}, {\methane} & LBL & LT  \\
ExoMol\tnm{b}       & {\water}, CO, {\carbdiox}, {\methane} & LBL & LT  \\
Partridge \& Schwenke & {\water} & LBL  & LT \\
Schwenke   & TiO & LBL  & LT  \\
Plez       & VO & LBL  & LT  \\
Borysow    & {\molhyd}--{\molhyd}, {\molhyd}--He & CS   & CIA \\
HITRAN\tnm{c}   & {\molhyd}--{\molhyd}, {\molhyd}--He & CS   & CIA \\
\hline
\end{tabular*}
\begin{minipage}{\linewidth}
\footnotesize {\bf Notes.}
\sp{a} HITRAN provides data for 47 additional species. \\
\sp{b} ExoMol provides data for 79 additional species. \\
\sp{c} and additional CIA pairs involving {\methane}, N$_2$, O$_2$, CO$_2$, Ar, and air (although we note that these databases have wavelength and temperature ranges significantly more limited than those for {\molhyd} and He).
\end{minipage}
\end{table}
}

\subsubsection{Cross-section Data}

Cross-section files provide tabulated opacities as a function of
temperature and wavenumber.  Since CIA varies smoothly across
wavenumber, this is the preferred format for CIA files.  Some
databases also provide line-transition data in cross-section format.

The {\transit} input cross-section files consist of ASCII tables of
the opacity normalized to the species number density (in units of
cm$\sp{-1}$amagat$\sp{-2}$, with $1\ {\rm amagat} = n\sb{0} =
2.68679\tttt{19}$ molecules cm$\sp{-3}$).

{\transit} provides Python scripts to format the CIA data files given
by the Borysow
group\footnote{\href{http://www.astro.ku.dk/~aborysow/programs}
{astro.ku.dk/{\sim}aborysow/programs}} \citep{BorysowEtal2001jqsrtH2H2highT,
Borysow2002jqsrtH2H2lowT, BorysowEtal1988apjH2HeRT,
BorysowEtal1989apjH2HeRVRT, BorysowFrommhold1989apjH2HeOvertones} and
HITRAN \citep{RichardEtal2012jqsrtCIA,
KarmanEtal2019icarHITRANupdateCIA} into its internal format.

\begin{figure}[tb]
\centering
\includegraphics[width=\linewidth, clip]{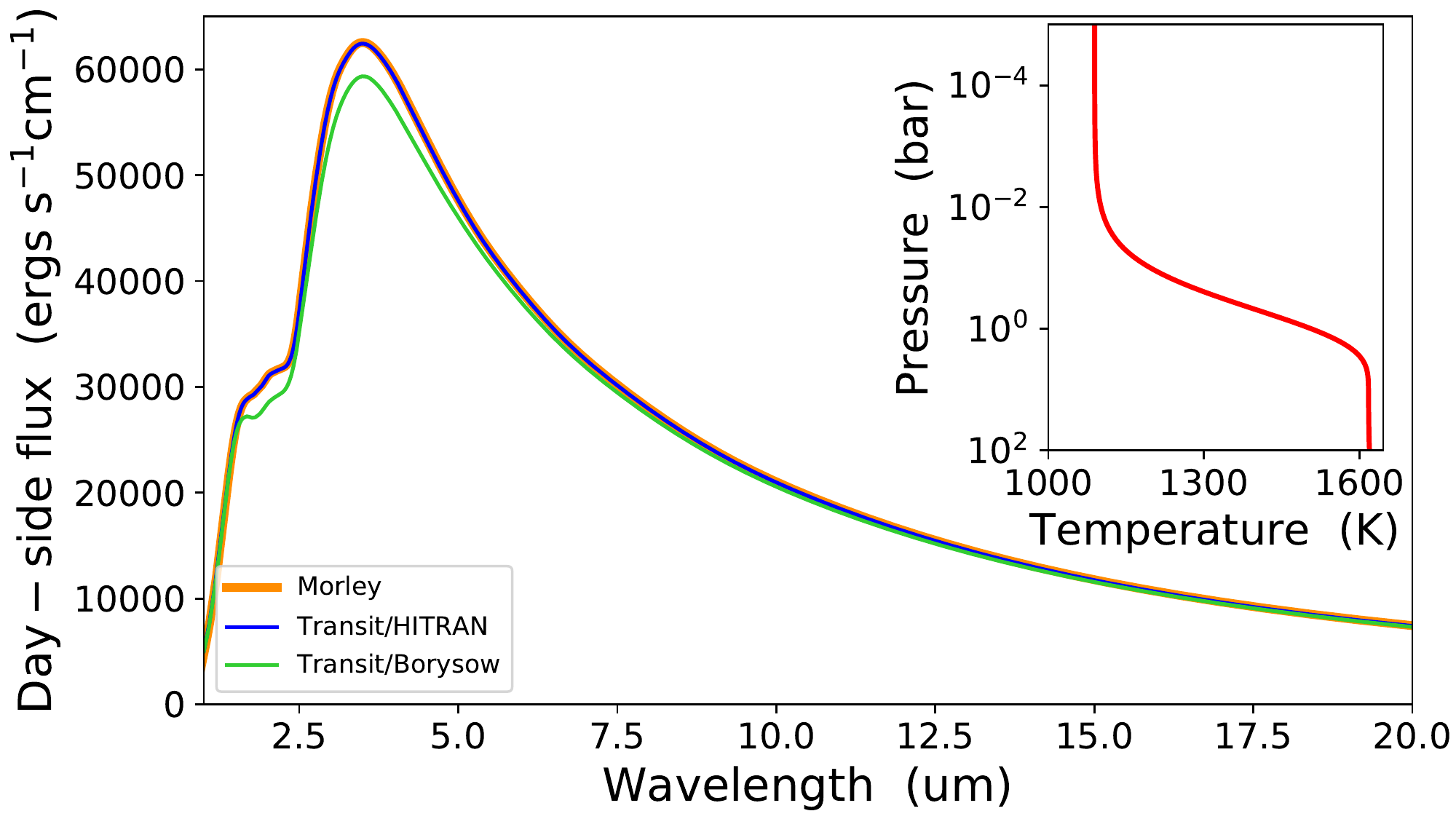}
\caption{Model emission spectra for pure collision-induced absorption.
  The blue and green curves show the {\transit} spectra for HITRAN and
  Borysow CIA opacities, respectively.  The orange curve shows the
  HITRAN CIA spectrum of C. Morley.  All spectra were calculated
  for an atmospheric model with uniform mole mixing ratios composed of
  85\% {\molhyd} and 15\% He, for a planet with a 1 {\rjup} radius and
  a surface gravity of 22 m\;s$\sp{-2}$.  The inset shows the
  atmospheric temperature profile as a function of pressure.}
\label{fig:CIAvalid}
\end{figure}

To evaluate the CIA opacity for a given atmospheric model, {\transit}
performs a bicubic interpolation (wavenumber and temperature) from the
tabulated CIA opacities ($e\sb{\rm cia}\sp{\rm t}$), and scales the
values to units of cm$\sp{-1}$ ($e\sb{\rm cia}$), multiplying by the
number density of the species:
\begin{equation}
e\sb{\rm cia} = e\sb{\rm cia}\sp{\rm t} \frac{n\sb{1}}{n\sb{0}}
                                        \frac{n\sb{2}}{n\sb{0}},
\end{equation}
where $n\sb{1}$ and $n\sb{2}$ are the number densities of the
colliding species (in units of molecules cm$\sp{-3}$).
Figure \ref{fig:CIAvalid} shows {\transit} emission spectra for pure
{\molhyd}--{\molhyd} and {\molhyd}--He CIA opacities.  When compared
against the models of C. Morley \citetext{priv.\ comm.}, we agree to
better than 0.5\%.  We also noted that the HITRAN absorption is weaker
than the Borysow absorption, producing emission spectra 2\%--8\%
stronger for this atmospheric model.

% SOURCE: /home/patricio/ast/esp01/bart/compendium/PhD_dissertation_Cubillos_UCF/011_Figure01_CIA.py

% Line-transition opacity:
{\transit} treats the cross-section line-transition opacity inputs in
a similar manner as the CIA data files, except that the opacity is
given in cm$\sp{-1}$amagat$\sp{-1}$ units.  {\transit} provides a
routine to re-format HITRAN and
ExoMol\footnote{ \href{http://www.exomol.com/data/data-types/xsec}
{http://exomol.com/data/data-types/xsec}} cross-section data files
into the format required by {\transit}.

\subsubsection{Line-by-line Data}

{\transit} stores the line-by-line opacity data into a Transition Line
Information (TLI) binary file.  A TLI file contains a header and the
line-transition data.  The header contains the number and names of
databases, species, and isotopes, and the partition function per
isotope as a function of temperature.  The line-transition data
consist of four arrays with the transition's wavelength, lower-state
energy, oscillator strength, and isotope ID.  The original {\transit}
line-reading code has been rewritten in Python to make it easier for
users to add functions to read additional line-list formats.
Currently, the {\transit} line reader can process line-transition
files from the HITEMP/HITRAN lists \citep{RothmanEtal2010jqsrtHITEMP,
GordonEtal2017jqsrtHITRAN2016}, the {\water} list
from \citet{PartridgeSchwenke1997jcpH2O}, the TiO list
from \citet{Schwenke1998fadiTiO}, and the VO list from B. Plez (priv.\
comm.).  Most line-by-line databases provide tabulated
partition-function files.  For the HITRAN and HITEMP databases,
{\transit} provides an implementation of the Total Internal Partition
Sums
code\footnote{\href{http://faculty.uml.edu/robert\_gamache/software/index.htm}
{http://faculty.uml.edu/robert\_gamache/software/index.htm}} \citep[][]{LaraiaEtal2011icarTIPS}. {\transit}
is also compatible with the open-source {\repack} line-compression
tool \citep{Cubillos2017apjRepack}, which allows it to process the
ExoMol line lists \citep{TennysonEtal2016jmsExomol}. By using
{\repack} to identify and retain only the strongest line transitions
that dominate the opacity spectrum of a given molecule, we improve the
performance of the radiative-transfer calculation by discarding the
large majority of weak transitions that have little impact on the
outputs.

\begin{figure}[tb]
\centering
\includegraphics[width=\linewidth]{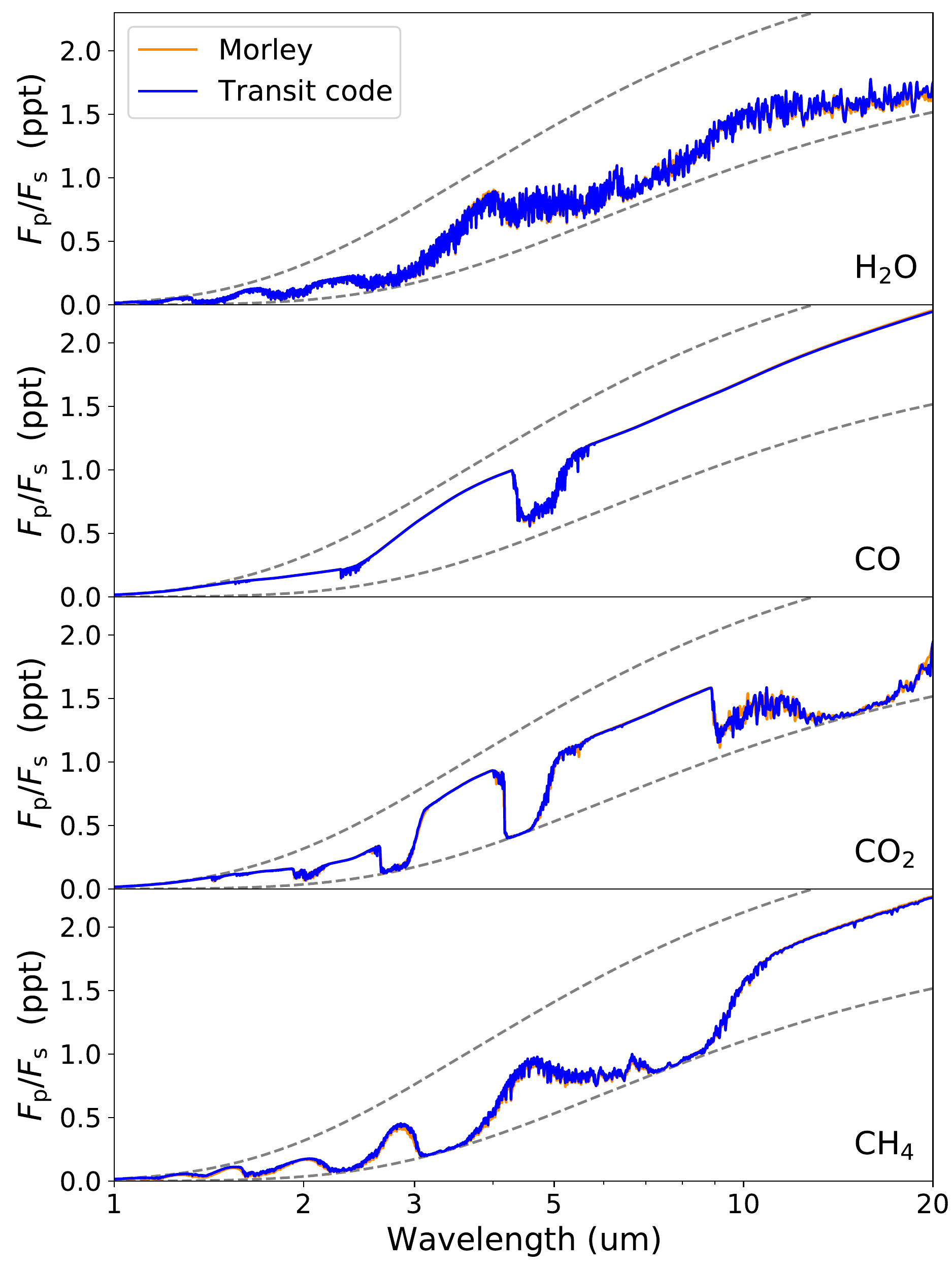}
\caption{Model planet-to-star flux ratio spectra for water,
  carbon monoxide, carbon dioxide, and methane (top to bottom panels).
  The blue and orange solid curves show the {\transit}-code and the
  Morley radiative-transfer spectra, respectively (Gaussian-smoothed
  for better visualization).  The planetary atmospheric model is the
  same as in Fig.\ (\ref{fig:CIAvalid}), with additional uniform
  mixing ratios of $\ttt{-4}$ for the respective species in each
  panel.  The stellar model corresponds to a blackbody spectrum of a 1
  {\rsun} radius and 5700 K surface-temperature star.  The dashed grey
  lines indicate the flux ratio for planetary blackbody spectra at
  1090 and 1620 K (atmospheric maximum and minimum temperatures).  The
  CIA opacity comes from \citet{RichardEtal2012jqsrtCIA}.  For
  {\water} and {\methane} both models used the line lists
  from \citep{PartridgeSchwenke1997jcpH2O}
  and \citet{YurchenkoTennyson2014mnrasExomolCH4}, respectively. For
  {\carbdiox}, Morley used \citet{HuangEtal2013jqsrtCO2}
  and \citet{HuangEtal2014jqsrtCO2}, whereas {\transit} used HITEMP.
  For CO, the Morley models used the line list
  from \citet{Goorvitch1994apjsCOlinelist}, whereas {\transit} used
  the HITEMP line list.  The spectrum for each species shows a good
  agreement between {\transit} and Morley's models.}
\label{fig:emission}
\end{figure}

\begin{figure}[tb]
\centering
\includegraphics[width=\linewidth, clip]{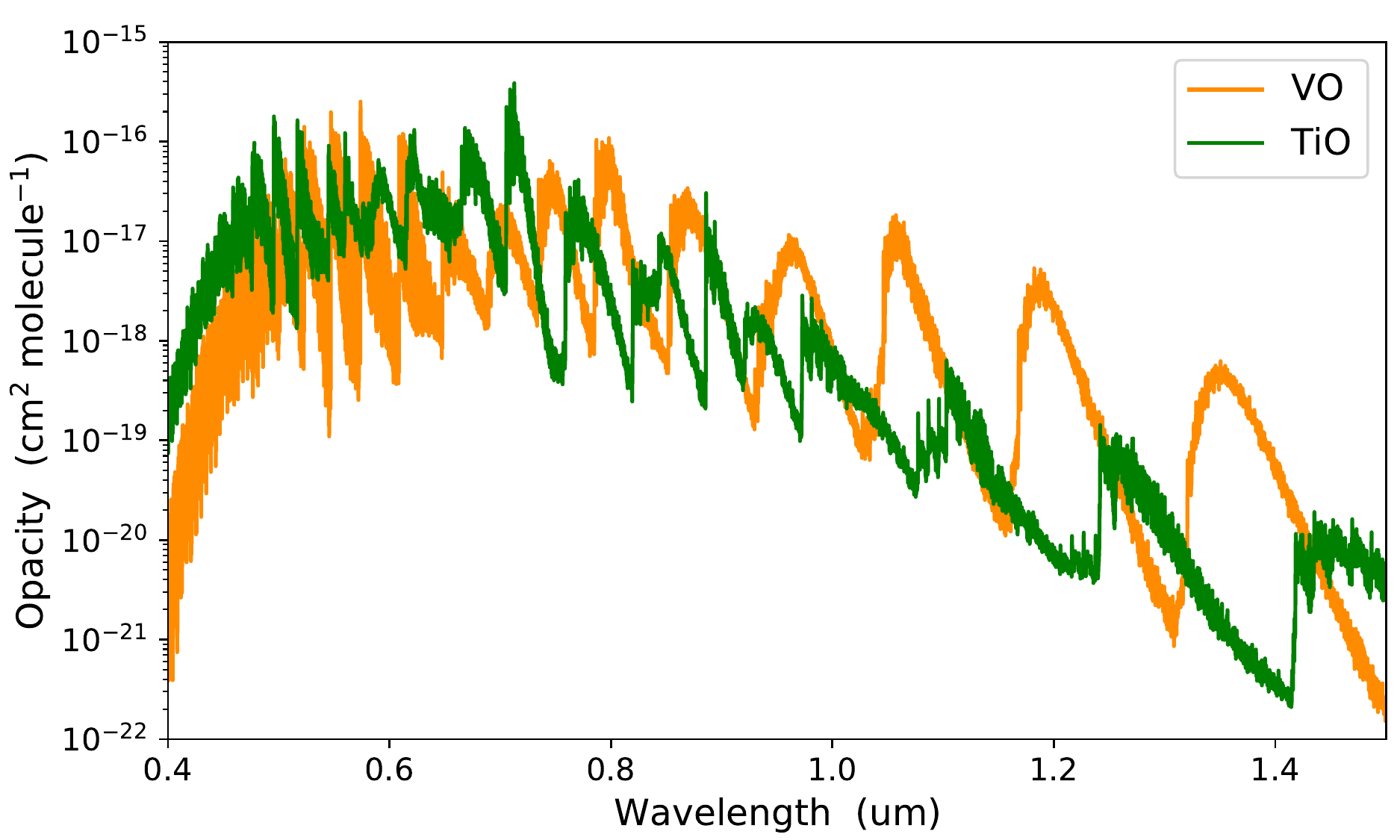}
\caption{Near-infrared titanium- and vanadium-oxide
  opacity spectra.  {\transit} calculated the opacities at a
  temperature of 2200 K and a pressure of 10.13 bar (10 atm).  Our
  models are consistent with Figures (4) and (5)
  of \citet{SharpBurrows2007apjOpacities}.}
\label{fig:TiO-VOopacity}
% SOURCE: /home/patricio/ast/esp01/bart/compendium/PhD_dissertation_Cubillos_UCF/04_Figure03_TiO-VO_opacity.py
\end{figure}

Figure \ref{fig:emission} shows an example of the emission spectra of
{\water}, CO, {\carbdiox}, and {\methane}.  A comparison with the
Morley models shows a good agreement for all four cases.
Figure \ref{fig:TiO-VOopacity} shows an example of the TiO and VO
opacity spectra.  The {\transit} spectra agree well with that
of \citet{SharpBurrows2007apjOpacities}.
{\citet{HarringtonEtal2021psjBART}} present additional tests of the
{\transit} and BART codes.

% SOURCE: /home/patricio/ast/esp01/bart/compendium/PhD_dissertation_Cubillos_UCF/012_Figure02_emission.py

\section{Application to HAT-P-11b}
\label{sec:analysis}

In this section we apply our atmospheric retrieval analysis to the
transit observations of the Neptune-sized planet HAT-P-11b, and
compare our results to previous analyses
by \citet[][]{FraineEtal2014natHATP11bH2O}
and \citet{ChachanEtal2019ajHATP11bPancet}.

The exoplanet HAT-P-11b \citep{BakosEtal2010apjHATP11b} is slightly
larger than Neptune in mass (26 $M\sb{\oplus}$) and radius (4.7
$R\sb{\oplus}$).  The planet orbits an active K4 dwarf star ($R\sb{\rm
s}=0.75$ {\rsun}, $T\sb{\rm s}= 4780$~K), at a distance of 0.053 AU,
with a period of 5 days.  Given these parameters, the planetary
equilibrium temperature (temperature at which the emission as
blackbody balances the absorbed energy, assuming zero albedo and
efficient heat redistribution) is $T\sb{\rm eq} = 878$ K.

{\citet{FraineEtal2014natHATP11bH2O}} observed five transits of
HAT-P-11b with the {\HubbleST}'s ({\HST}) Wide Field Camera 3 (WFC3)
G141 grism, covering the 1.1 to 1.7~{\micron} region of the spectrum,
and the {\Spitzer} Infrared Array Camera (IRAC) 3.6 and 4.5~{\microns}
bands.  \citet{ChachanEtal2019ajHATP11bPancet} reported eight
additional transits observed with {\HST}\/'s STIS (G403L and G750L)
and WFC3 (G102), providing a continuous wavelength coverage from 0.34
to 1.1 {\micron}.  \citet{FraineEtal2014natHATP11bH2O}
and \citet{ChachanEtal2019ajHATP11bPancet} presented transmission
spectra binned in increasing resolving power ($R\sim$8, 15, 42, 46)
with increasing wavelength (G430L, G750L, G102, G141, respectively).

The transmission spectrum of HAT-P-11b is relatively featureless:
there is no evidence for sodium (0.59 {\micron}) nor potassium
absorption (0.77 {\micron}), and there is an evident but muted {\water}
absorption feature (1.4 {\micron}) compared to that of an atmosphere
with solar-abundance composition.  Noteworthy,
both \citet{FraineEtal2014natHATP11bH2O}
and \citet{ChachanEtal2019ajHATP11bPancet} consistently find that the
{\Spitzer} 3.6- and 4.5-{\micron} transit depths are relatively
shallow compared to the WFC3 transit depths.  We re-analyzed the
{\Spitzer} light curves finding similar results
(Appendix \ref{sec:poet}).  \citet{FraineEtal2014natHATP11bH2O}
attributed this anomaly to stellar brightness variations between the
different epochs, and estimated a transit depth offset of $\sim
93$~ppm between {\HST}\//G141 and {\Spitzer}.
However, \citet{ChachanEtal2019ajHATP11bPancet} concluded from their
photometric monitoring of the stellar spot coverage that such a large
offset cannot be explained by stellar activity alone.  In principle,
given the separation in time between the {\Spitzer} observations, the
stellar variability should inflict a flux offset {\Spitzer}
observations; however, \citet{FraineEtal2014natHATP11bH2O} found that
the stellar-spot variability had a negligible impact at these
wavelengths.

{\citet{FraineEtal2014natHATP11bH2O}} characterized the atmospheric
composition with the Self-Consistent Atmospheric Retrieval Framework
for Exoplanets (SCARLET) tool \citep{
BennekeSeager2012apjRetrieval,
BennekeSeager2013apjRetrievalClouds}
% Benneke2015arxivSCARLETretrieval
%, BennekeEtal2019natasGJ3470bMieClouds
whereas \citet{ChachanEtal2019ajHATP11bPancet} used
the \textsc{PLATON} atmospheric retrieval
framework \citep{ZhangEtal2019paspPLATON}.

{\renewcommand{\arraystretch}{1.05}
\begin{table*}[tb]
\centering
\caption{BART Atmospheric Retrievals of HAT-P-11b}
\label{table:retrieval}
\begin{tabular*}{\linewidth} {@{\extracolsep{\fill}} lcccc}
\hline
\hline
& Analyzed dataset:         & \citet{FraineEtal2014natHATP11bH2O}
    & \citet{ChachanEtal2019ajHATP11bPancet}  & \citet{ChachanEtal2019ajHATP11bPancet}  \\
Parameter                      & Prior  &  {\HST}\//G141 + {\Spitzer}\/\tnm{a} & {\HST}\/\tnm{a} & {\HST} + {\Spitzer}\/\tnm{a} \\ 
\hline
$T_0$ (K)                          & Uniform (200, 3000) &  $860_{-330}^{+420}$  &    $1120_{-210}^{+245}$  &  $1120_{-140}^{+180}$    \\
$R_p (R_\oplus)$ at $p_0=0.1$ bar  & Uniform (2.0, 7.0)  &  $4.19_{-0.19}^{+0.08}$ &  $4.12_{-0.13}^{+0.17}$ & $4.05_{-0.09}^{+0.09}$   \\
$\log_{10}(p_{\rm cloud}/{\rm bar})$ & Uniform (-6, 2)   &  $-1.9_{-1.9}^{+2.4}$  &   $-2.8_{-0.9}^{+1.8}$  &  $-3.6_{-0.5}^{+0.9}$    \\
$\log_{10}(\kappa_{\rm ray})$      & Uniform (-5, 5)     &  $\cdots$           &  $-1.2_{-2.5}^{+2.4}$  &  $-0.7_{-3.0}^{+2.6}$     \\
{\HST}--{\Spitzer} offset (ppm)   & Uniform (-200, 200) &   $102_{-27}^{+34}$  & $\cdots$               &  $\cdots$              \\
$\log_{10}({\rm H2O})$             & Uniform (-10, 0)    &  $-1.5_{-1.5}^{+0.8}$  &   $-2.9_{-1.7}^{+1.3}$  &  $-1.7_{-1.0}^{+0.8}$    \\
$\log_{10}({\rm CH4})$             & Uniform (-10, 0)    &  $-6.7_{-2.2}^{+2.3}$  &   $-2.6_{-1.7}^{+1.4}$  &  $-7.5_{-1.7}^{+1.8}$    \\
$\log_{10}({\rm CO})$              & Uniform (-10, 0)    &  $-5.4_{-3.1}^{+3.1}$  &   $-5.6_{-3.0}^{+3.1}$  &  $-6.4_{-2.5}^{+2.7}$    \\
$\log_{10}({\rm CO2})$             & Uniform (-10, 0)    &  $-6.7_{-2.2}^{+2.6}$  &   $-5.3_{-3.2}^{+2.7}$  &  $-7.8_{-1.5}^{+1.8}$    \\
\hline
\end{tabular*}
\begin{minipage}{\linewidth}
\footnotesize {\bf Notes.} \sp{a} Reported values correspond to the
marginal posterior distribution's median and boundaries of the 68\%
central credible interval \citep{Andrae2010arxivErrorEstimation}. \\
\end{minipage}
\end{table*}
}

\subsection{Atmospheric Retrieval}
\label{sec:retrieval}

We applied our atmospheric retrieval analysis under three different
scenarios.  First, we retrieved on the transmission spectrum
of \citet{FraineEtal2014natHATP11bH2O}, i.e., considering only the
{\HST}\//WFC3 G141 and {\Spitzer}, including a free parameter to model
an offset between the {\HST} and {\Spitzer} transit depths.  In the
second and third scenarios, we modeled the transmission spectrum
of \citet{ChachanEtal2019ajHATP11bPancet} with and without the
{\Spitzer} observations.  For these latter scenarios, we used the
stellar-activity-corrected spectra (thus, we do not include an
transit-depth offset free parameter).

We used the {\BART} package to retrieve and constrain the atmospheric
temperature, composition, gray cloud deck, and Rayleigh scattering
absorption.  We modeled the planetary atmosphere with a set of 101
layers, equi-spaced in log-pressure, ranging from $\ttt{2}$ to
$\ttt{-8}$ bar.  The retrieval considers the optical and near-infrared
spectrum between 0.34 and 5.5 {\microns}.
Following \citet{ChachanEtal2019ajHATP11bPancet}, we model the
atmospheric temperature as an isothermal profile.  Unlike
both \citet{FraineEtal2014natHATP11bH2O}
and \citet{ChachanEtal2019ajHATP11bPancet}, we did not assume
abundances in thermochemical equilibrium, since this is highly
unlikely at the temperatures expected for HAT-P-11b, as disequilibrium
processes start to take place at $T\lesssim
2000$~K \citep[][]{Moses2014rsptaChemicalKinetics}.  We thus modeled
the composition as constant-with-altitude volume-mixing-ratios
profiles for {\water}, {\methane}, CO, and {\carbdiox} (the main
species expected to dominate the observed transmission spectrum).
While the constant-with-altitude assumption is likely unphysical, from
a statistical standpoint, the limited precision and spectral coverage
of current observations does not allow to constrain more complex
models.  Thus, the adoption of such simplistic retrieval
approximations should be considered as first-order estimations.
Certainly, the next generation of telescopes will require a more
accurate description of the planetary physical
properties \citep[e.g.,][]{RocchettoEtal2016apjJWSTbiases,
BlecicEtal2017apj3Dretrieval, CaldasEtal2019aaTransmission3Deffects}.
The code assumes that the planet has a primary atmosphere, where the
remaining bulk of the composition is assumed to be {\molhyd} and He at
a solar-composition ratio \citep{AsplundEtal2009araSolarComposition},
such that the total mixing ratio is 1.0.  We neglect the contribution
by Na and K, since the data does not indicate that they are
detectable.  We used the HITEMP opacities for
{\methane} \citep{HargreavesEtal2020apjsHitempCH4},
CO \citep{LiEtal2015apjsCOlineList}, and
{\carbdiox} \citep{RothmanEtal2010jqsrtHITEMP}.  For {\water} we used
the ExoMol opacities \citep{PolyanskyEtal2018mnrasPOKAZATELexomolH2O}
processed with the {\repack} line compression
tool \citep{Cubillos2017apjRepack}.  The model also included CIA
opacities for {\molhyd}-{\molhyd} and
{\molhyd}-He \citep{KarmanEtal2019icarHITRANupdateCIA}.  For the
calculation of the line-by-line opacities, we set a cutoff at
$\ttt{-10}$ times the strongest lines.  This threshold cutoff can
introduce transit-depth biases on the order of $\lesssim 10^{-1}$~ppm
(far lower than the typical observing uncertainty), while
significantly speeding up the calculations. We also considered a gray
cloud deck in the atmospheric model parameterized by the cloud-deck
top pressure $p_{\rm cloud}$.  For the retrievals of
the \citet{ChachanEtal2019ajHATP11bPancet} spectrum we also included
the parametric Rayleigh-scattering model described in
Section \ref{sec:clouds}.

{\BART} calculates the altitude (the radius) of each layer
using the hydrostatic-equilibrium equation:
\begin{equation}
\frac{\der p}{\der z} = -\rho g = -\rho(z)\frac{G M_{\rm p}}{z^2},
\end{equation}
where $z$ and $\rho$ are the altitude and mass density of each layers,
respectively, and $g(z)=GM_{\rm p}/z^2$ is the gravity, with $G$ the
gravitational constant and $M_{\rm p}$ the mass of the planet.  A
retrieval free parameter sets the reference planetary radius $R_{\rm
p}$ at a fiducial pressure level of 0.1~bar.

One of the main differences in the modeling approach
between \citet{FraineEtal2014natHATP11bH2O}
and \citet{ChachanEtal2019ajHATP11bPancet} is how they combine data
from non-simultaneous observations.
While \citet{FraineEtal2014natHATP11bH2O} includes a free parameter
that offsets the WFC3 and {\Spitzer} data
points, \citet{ChachanEtal2019ajHATP11bPancet} apply a
wavelength-dependent stellar-activity correction leading to shallower
transit depths at short wavelengths and larger transit depths at
longer wavelengths.  In this study, we aimed at reproducing the main
analyses of \citet{FraineEtal2014natHATP11bH2O}
and \citet{ChachanEtal2019ajHATP11bPancet}, and thus we adopted these
same assumptions.
%%%
Table \ref{table:retrieval} summarizes the retrieval free parameters
and their priors for each of the three scenarios.

\begin{figure*}[t]
\centering
\includegraphics[width=0.8\linewidth, clip]{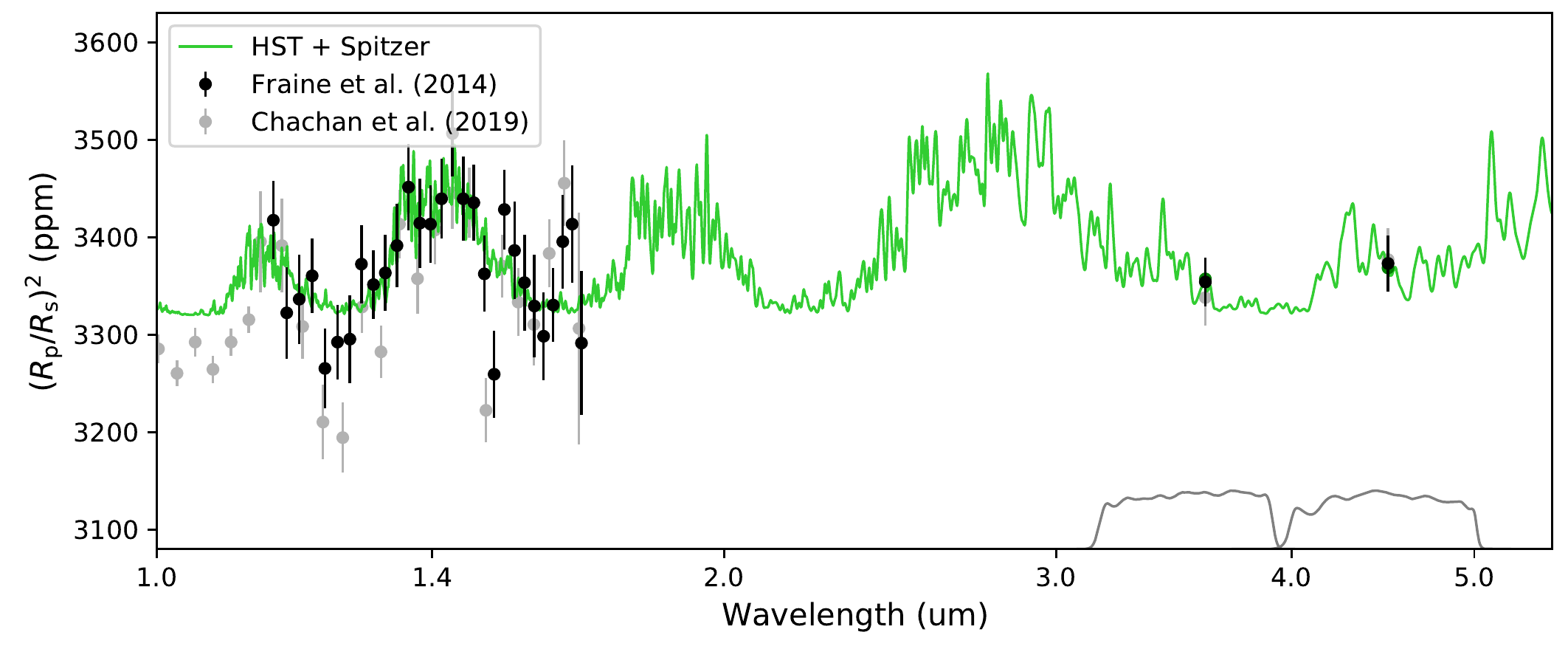}
\includegraphics[width=0.8\linewidth, clip]{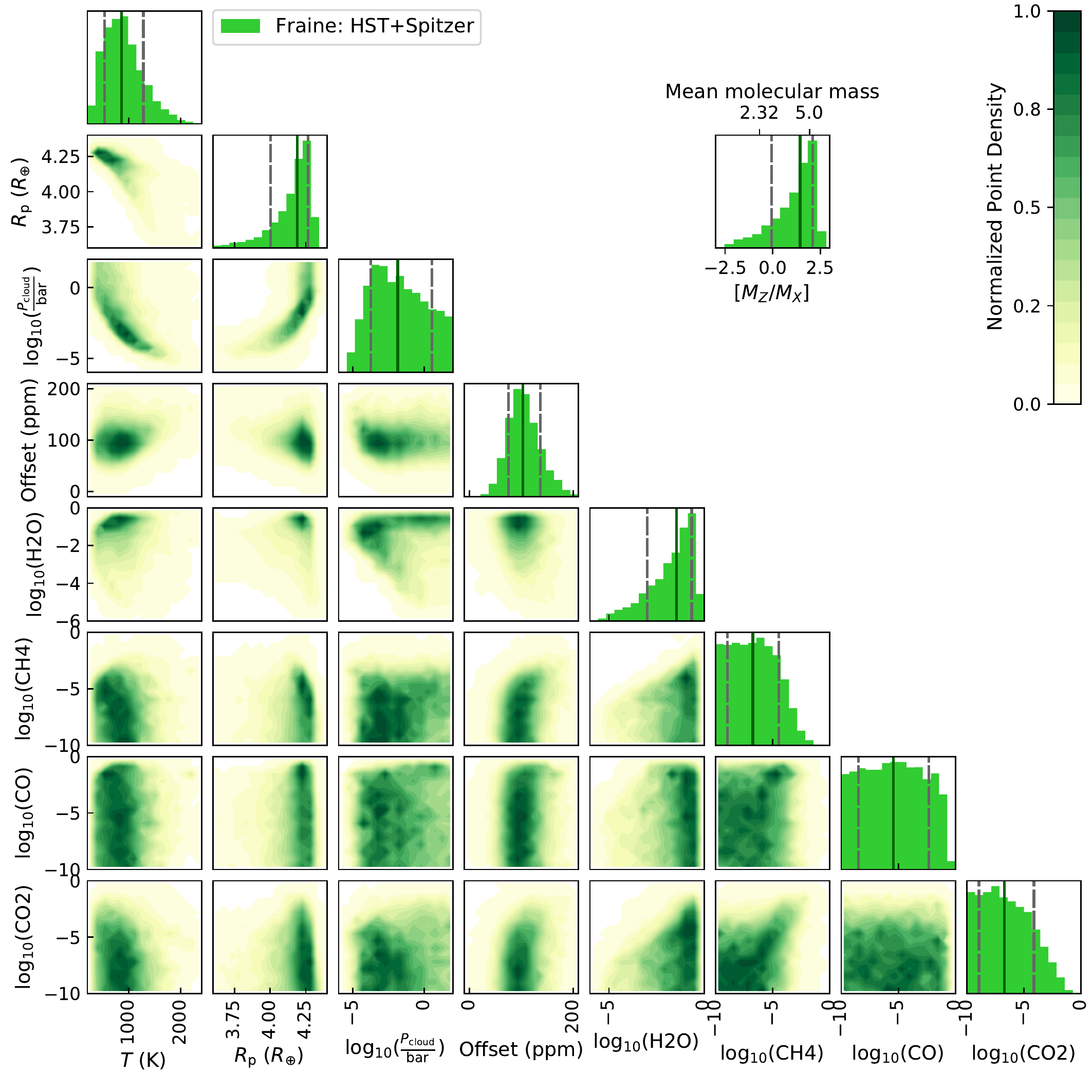}
\caption{
  {\bf Top:} HAT-P-11b transmission spectra.  The black points with
  error bars denote the {\HST} WFC3/G141 and {\Spitzer} IRAC data and
  their $1\sigma$ uncertainties reported
  by \citet{FraineEtal2014natHATP11bH2O}.  The gray points with error
  bars denote the data reported
  by \citet{ChachanEtal2019ajHATP11bPancet}.  The green curve shows
  the {\BART} best-fitting spectrum to
  the \citet{FraineEtal2014natHATP11bH2O} dataset.  The {\HST} points
  were adjusted downwards according to the retrieved offset.  The gray
  curves at the bottom show the {\Spitzer} transmission filters.  {\bf
  Bottom:} Pairwise and marginal posterior probability distribution to
  the \citet{FraineEtal2014natHATP11bH2O} dataset.  The solid and
  dashed vertical lines denote the posterior's median and boundaries
  of the 68\% credible interval, respectively
  (Table \ref{table:retrieval}).}
\label{fig:bart_fraine}
% SOURCE: /home/pcubillos/ast/compendia/CubillosEtal2020_HATP11b/figure_HATP11b_retrieval_fraine.py
\end{figure*}

\begin{figure*}[p]
\centering
\includegraphics[width=0.8\linewidth, clip]{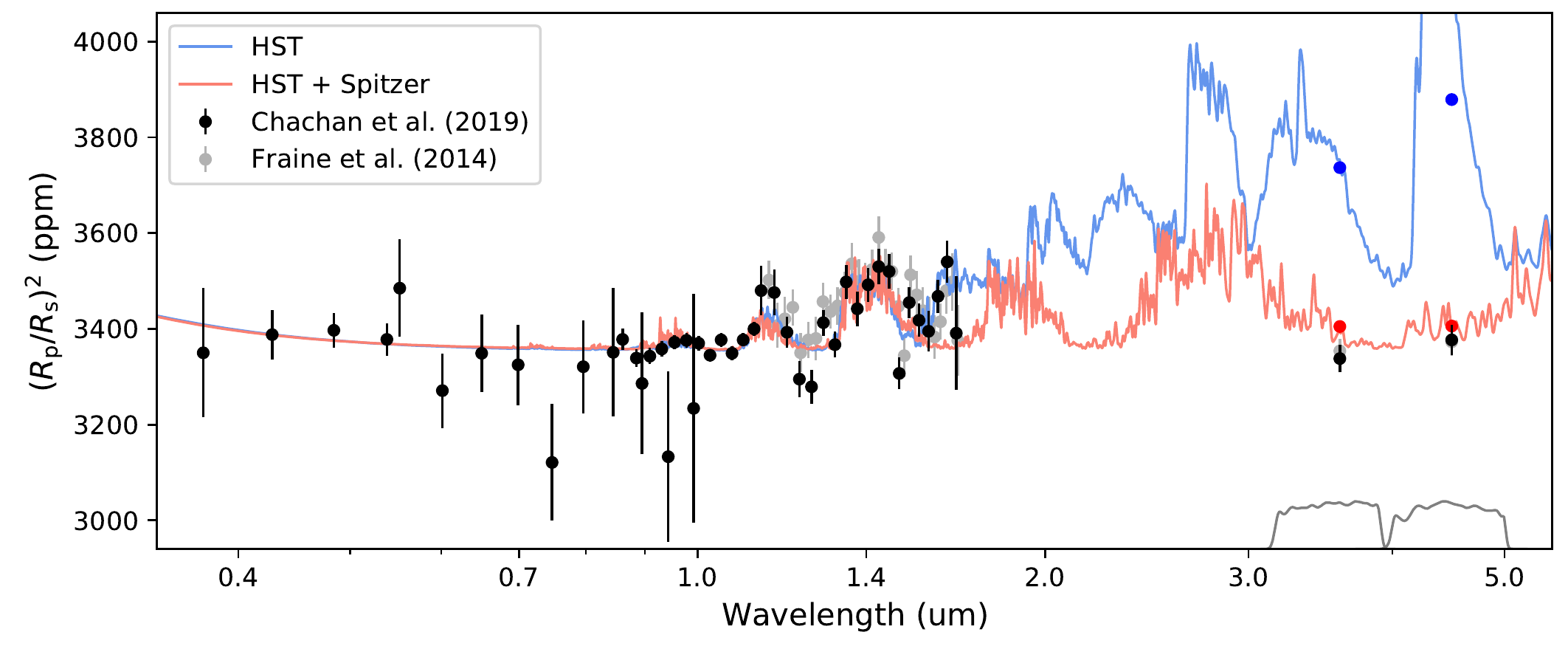}
\includegraphics[width=0.8\linewidth, clip]{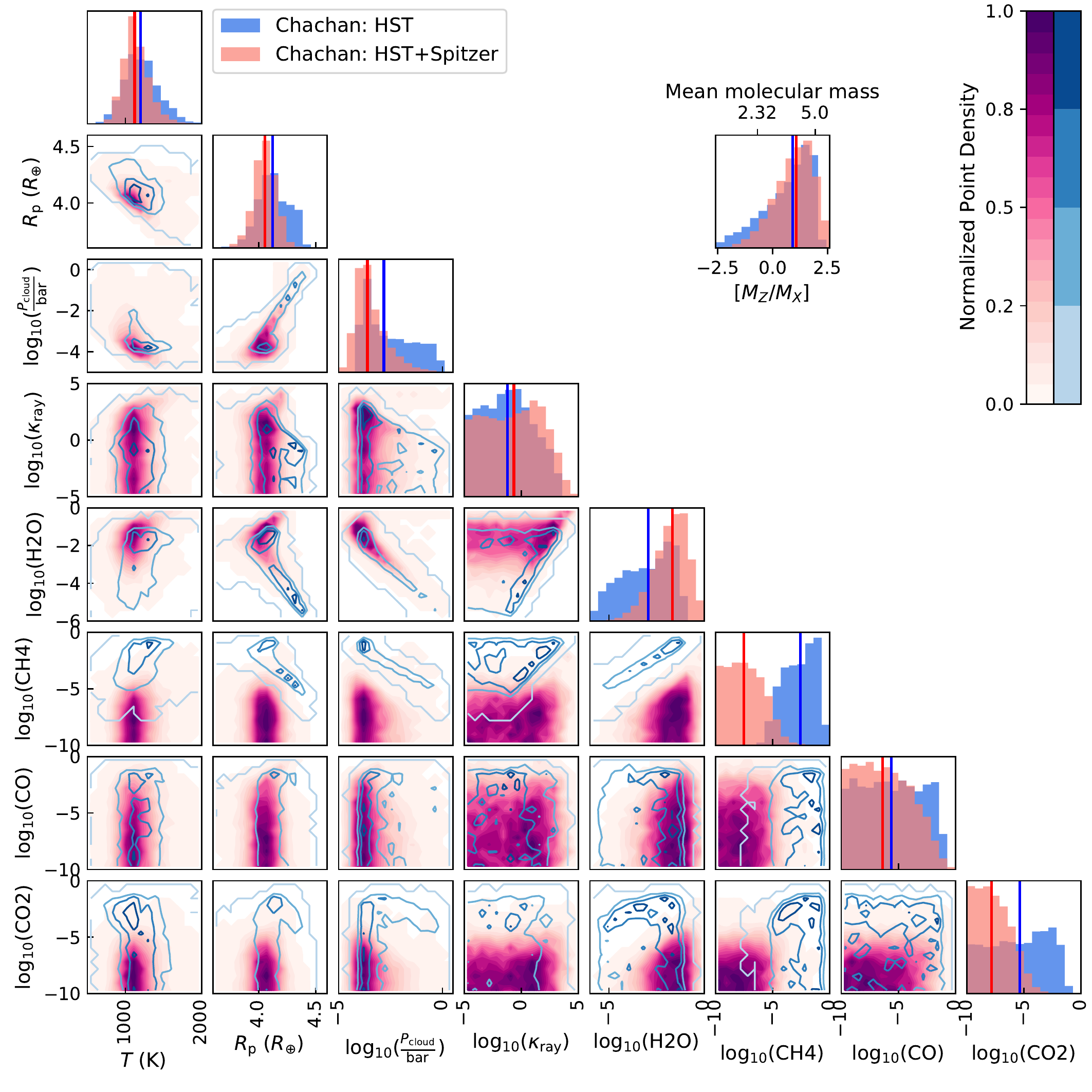}
\caption{ {\bf Top:} HAT-P-11b transmission spectra.  The black points
  with error bars denote the {\HST} and {\Spitzer} IRAC data and their
  $1\sigma$ uncertainties reported by
  \citet{ChachanEtal2019ajHATP11bPancet}.  The gray points with error
  bars denote the data reported by
  \citet{FraineEtal2014natHATP11bH2O}.  The pink and blue curves show
  the {\BART} best-fitting spectrum to the
  \citet{ChachanEtal2019ajHATP11bPancet} datasets (with and without
  considering the {\Spitzer} observations, respectively).  The gray
  curves at the bottom show the {\Spitzer} transmission filters.  {\bf
    Bottom:} Pairwise and marginal posterior probability distribution
  to the \citet{ChachanEtal2019ajHATP11bPancet} dataset (the
    pink shaded areas show the retrieval posteriors of the {\HST} and
    {\Spitzer} data, whereas the blue contours show that of the {\HST}
    data alone). The red and blue solid vertical lines denote the
  posterior's median for the retrieval with and without considering
  the {\Spitzer} observations, respectively.  We omitted plotting the
  credible interval boundaries to avoid cluttering (their values are
  available in Table \ref{table:retrieval}).}
\label{fig:bart_chachan}
% SOURCE: /home/pcubillos/ast/compendia/CubillosEtal2020_HATP11b/figure_HATP11b_retrieval_fraine.py
\end{figure*}

\subsubsection{\citet{FraineEtal2014natHATP11bH2O} Retrieval Results}
\label{sec:retresults1}

Figure \ref{fig:bart_fraine} and Table \ref{table:retrieval} show the
best-fitting spectrum and posterior distributions for the BART
atmospheric retrieval on the \citet{FraineEtal2014natHATP11bH2O} data
set.  Our best-fitting model yields $\chi^2_{\rm red}=1.05$,
indicating a good fit.  Although the comparison is not direct, we
found results that are qualitatively similar to those
of \citet{FraineEtal2014natHATP11bH2O}.  We found an offset between
{\HST} and {\Spitzer} with a median of $103_{-28}^{+34}$ ppm,
consistent with the 93~ppm value reported
by \citet{FraineEtal2014natHATP11bH2O}.
%%%
The pairwise posterior distributions show little correlation between
the offset and the other parameters; however, as discussed
by \citet{YipEtal2021ajGroundAndSpaceWASP96b}, such treatments are
often {\it ad hoc} corrections, and the results should be taken with
care.
%%%
The $\log({\rm H_2O})$--$\log(p_{\rm cloud})$ pairwise panel shows two
solution modes: the atmosphere can either have a high super-solar
{\water} abundance independent of the cloud coverage or an
anti-correlated {\water}--$p_{\rm cloud}$ mode.  This is qualitatively
consistent with the posterior distribution shown in Fig.~3
of \citet{FraineEtal2014natHATP11bH2O}.  Our {\water} marginal
posterior peaks at $\sim${}$250\times$ solar metallicity (the solar
abundance of {\water} in thermochemical equilibrium at 900~K is
$\sim$$\ttt{-3}$).  The posterior distribution of the
cloud-top pressure is broad, constrained to values $p_{\rm
cloud}\gtrsim 20$ mbar.  Both results are consistent with those
of \citet{FraineEtal2014natHATP11bH2O}, who found a best-fit
metallicity of $190\times$ solar and constrained the cloud deck to
pressures higher than $10$~mbar.  From our posterior we derived a
super-solar metals mass fraction of $[M_Z/M_X] = 1.4_{-1.5}^{+0.7}$,
with the caveat that we estimated this value from the molecular
abundances (largely dominated by {\water}),
whereas \citet{FraineEtal2014natHATP11bH2O} directly fit for the
metallicity.

The key to constrain the enhanced atmospheric abundance is the
amplitude of the 1--2 {\micron} absorption feature.  A higher
abundance of heavy elements implies a higher mean molecular mass,
hence a smaller scale height, and a smaller feature amplitude.  The
high {\water} abundance, thus, allows the model to match the observed
{\water}-band amplitude between 1 and 2 {\microns}.  The cloudy
solution results from the combination of the cloud deck, {\water}
abundance, and hydrostatic-equilibrium solution.  As the {\water}
abundance increases, the amplitude of the 1.4~{\micron} feature
becomes more prominent.  The presence of the gray cloud deck mutes the
amplitude of the feature to maintain the match to the observed
amplitude (hence the anti-correlation between $p_{\rm cloud}$ and
{\water} abundance).  At the same time, the $T$ and $R_{\rm p}$ at 0.1
bar must also vary to preserve the absolute value of the transit depth
at the observed values (see the correlation in the pairwise posteriors
in Fig.\ \ref{fig:bart_fraine}).

The abundances of CO, {\carbdiox}, and {\methane} remained largely
unconstrained, showing nearly flat posterior histograms, finding at
most upper limits.  This is an expected outcome, since the {\water}
opacity dominates the entire spectral range probed by these
observations, whereas the CO and {\carbdiox} abundance hinge
predominantly on the {\Spitzer} 4.5 {\micron} observation and the
{\methane} abundance hinges predominantly on the {\Spitzer} 3.6
{\micron} observation.  Combined with the larger number and better
spectral resolution of the {\HST} observations, the given
low-resolution data can better constrain the {\water} features at the
shorter wavelengths, whereas CO, {\carbdiox}, and {\methane} can only
be constrained if their spectroscopic signatures raise above the level
set by the {\water} absorption at the longer wavelengths.

We observed no artifacts in the posterior sampling due to the discrete
nature of the atmospheric profile model.  This is a particularly
relevant concern when including cloud models that make the atmosphere
instantly opaque below a certain pressure level, because the resulting
cloud deck location gets discretized by the atmospheric pressure
sampling.  In this case, it is likely that the posterior distributions
constrained by these datasets are broad enough such that effect of the
discrete sampling are not seen (e.g., over 1 dex for $p_{\rm
cloud}$).
\citet{CubillosBlecic2021mnrasPyratBay} show a better
implementation of a cloud-deck model by interpolating the atmospheric
sampling model at the pressure location of the opaque cloud deck.

\subsubsection{\citet{ChachanEtal2019ajHATP11bPancet} Retrieval Results}
\label{sec:retresults1}

Figure \ref{fig:bart_chachan} and Table \ref{table:retrieval} show the
best-fitting spectra and posterior distributions for the BART
atmospheric retrievals on the \citet{ChachanEtal2019ajHATP11bPancet}
data sets.  We found reasonably good fits to the {\HST} observations;
however, no model were able to fit simultaneously the transit depths
of the {\HST} and {\Spitzer} observations.  Given the {\HST}
observations, all models predict larger transit depth at the
{\Spitzer} bands.  Consequently, our best-fitting $\chi^2_{\rm red}$
values increased from 1.5 to 2.3 when we include the {\Spitzer} data
in the retrieval.  \citet{ChachanEtal2019ajHATP11bPancet} found
similar results, with $\chi^2_{\rm red}$ increasing from 1.9 to 2.8.
Since the retrievals of \citet{ChachanEtal2019ajHATP11bPancet}
included a free parameter scaling the instrumental errors
($\sigma_{\rm mult}$), they naturally obtained a larger $\sigma_{\rm
mult}$ when including the {\Spitzer} observations, and thus found
broader posterior distributions.  In contrast, we generally found
narrower posterior distributions when including the {\Spitzer}
observations.  As expected, when the retrieval did not consider the
{\Spitzer} observations, the CO and {\carbdiox} abundances are
completely unconstrained (the dominant absorption feature of these
molecules is in the 4.5 {\micron} band).  Including the shallow
{\Spitzer} transit depths places more stringent constraints on the CO
and {\carbdiox} abundance upper limits.

The {\methane} abundance posterior is an interesting case,
the \citet{ChachanEtal2019ajHATP11bPancet} {\HST} transmission
spectrum shows an increase with wavelength at the red-end tail
($\sim$1.6 {\microns}) that coincides with a {\methane} band.  Thus,
the {\HST} data alone suggest a high {\methane} abundance ($\log({\rm
CH_4})=-2.6_{-1.7}^{+1.4}$), which is discouraged by the inclusion of
the shallow 3.6 {\micron} observation ($\log({\rm
CH_4})=-7.5_{-1.7}^{+1.8}$).  In fact, the retrieval on the {\HST}
data alone places a tight correlation between the {\water} and
{\methane} abundances (with {\methane} slightly more abundant) that
extends over several orders of magnitude.  Including the {\Spitzer}
data breaks the {\water}--{\methane} correlation, allowing the
{\water} abundance to adopt higher values \citep[in a similar fashion
to the run without {\methane} of][]{ChachanEtal2019ajHATP11bPancet}.

Overall, we found similar behavior for the retrieved haze and cloud
parameters as \citet{ChachanEtal2019ajHATP11bPancet}.  The cloud-top
pressures adopted lower values when including the {\Spitzer}
observations, though we obtained $p_{\rm cloud}$ values $\sim$1.5 dex
lower than those of \citet{ChachanEtal2019ajHATP11bPancet}.  The
Rayleigh scattering scale factor remained largely unconstrained for
both scenarios, with upper limits $\sim$1 dex larger than the expected
values for {\molhyd} Rayleigh scattering.

We derived super-solar metals mass fraction (on the order of
$10\times$ solar) for both the {\HST}-only retrieval ($[M_Z/M_X] =
0.9_{-1.5}^{+0.8}$) and the retrieval including the {\Spitzer}
observations ($[M_Z/M_X] = 1.1_{-1.0}^{+0.8}$). In
contrast, \citet{ChachanEtal2019ajHATP11bPancet} found $10\times$
sub-solar and $100\times$ super-solar metallicities, respectively.

We do not have a clear explanation for this discrepancy.  We can only
speculate that this occurs because of the difference between
thermochemical-equilibrium chemistry
\citep{ChachanEtal2019ajHATP11bPancet} and free constant-with-altitude
chemistry (this work), which is arguably the main difference between
these two analyses.  An {\it a posteriori}
  equilibrium-chemistry calculation according to our best-fit results
  ($T=1120$~K and $1.1\times$ solar metallicity) shows CO/{\water}
  ratios near unity, inconsistent with our posterior distributions,
  which show lower relative CO abundances.  It is likely that the set
of high-metallicity solutions found here are not allowed under
chemical equilibrium (at least under the assumption of solar
  elemental ratios), thus leading to different posterior
distributions.  We note also that equilibrium chemistry often
  leads to non constant-with-altitude abundace profiles, particularly
  for non-isothermal temperature profiles, which makes the discrepancy
  between our results and those of
  \citet{ChachanEtal2019ajHATP11bPancet} less simple to interpret.
Finally, we note that the correlations between parameters a strongly enhanced
when we neglect the {\Spitzer} observations (Fig.\
\ref{fig:bart_chachan}).

\section{Conclusions}
\label{sec:conclusions}

In this article we presented the open-source, open-development
{\transit} code to compute exoplanet spectra.  {\transit} is a core
package of the {\BART} project to characterize exoplanet atmospheres
in a statistically robust manner.  {\BART} is jointly presented here,
in \citet{HarringtonEtal2021psjBART}, and
in \citet{BlecicEtal2021psjBART}.  The {\transit} package solves the
one-dimensional radiative-transfer equation to calculate transmission
or emission spectra.  The current version of the code accounts for
opacities from molecular line-transition, collision-induced
absorption, gray cloud decks, and Rayleigh scattering.  We compared
the {\transit} emission spectra against models of C. Morley, finding a
good agreement.  {\transit} also accurately reproduced the opacity
spectra of \citet{SharpBurrows2007apjOpacities}.  The project's source
code and documentation are available
at \href{https://github.com/exosports/transit}
{https://github.com/exosports/transit}.

We further applied the {\BART} atmospheric retrieval analysis to the
{\HST} and {\Spitzer} transit observations of
HAT-P-11b \citep{FraineEtal2014natHATP11bH2O,
ChachanEtal2019ajHATP11bPancet} assuming an isothermal temperature
profile, constant-with-altitude composition of {\water}, CO,
{\carbdiox}, and {\methane} (i.e., without assuming equilibrium
chemistry), gray clouds, and Rayleigh scattering.  The transmission
spectra constrains mainly the {\water} abundance through its 1.4
{\micron} band.  When retrieving on the transmission spectra
of \citet{FraineEtal2014natHATP11bH2O} alone ({\HST\// WFC3/G141} and
{\Spitzer}) we found a $\sim$10--100$\times$ super-solar enhancement
of metals, determined by the retrieved high {\water} abundance.
Alternatively, models with high-altitude cloud decks fit the
observations well.  Other species abundances ({\methane}, CO, and
{\carbdiox}) remained largely unconstrained.

Given the significant stellar activity of HAT-P-11, it is likely that
the {\HST} and {\Spitzer} transit depths have an offset due to the
unknown absolute calibration of the system flux.  These retrievals
made use of a free parameter to account for this offset; however, this
extra degree of freedom can mask spectral features, limiting our
capacity to find atmospheric constraints.  These results are
consistent with the atmospheric retrieval analysis
of \citet{FraineEtal2014natHATP11bH2O}.

When retrieving on the transmission spectra
of \citet{ChachanEtal2019ajHATP11bPancet}, our result do not totally
agree with the previous work.  {\citet{ChachanEtal2019ajHATP11bPancet}
concluded that their atmospheric analysis is primarily driven by the
{\HST} observations.}  However, when taking the transmission data at
face value, the {\Spitzer} observations change the outputs
significantly.  Neither their nor our models were able to fit well all
observations simultaneously.  By including the {\Spitzer}
observations, the \citet{ChachanEtal2019ajHATP11bPancet} retrievals
favored metallicities three orders of magnitude larger.  In our case,
the shallow {\Spitzer} 3.6 {\micron} transit depth (relative to the
{\HST} transit depth) ruled out the presence of {\methane} in the
atmosphere, in contrast to the {\methane} abundances higher than
{\water} for the retrieval without the {\Spitzer} observations.
Though, for both of our retrieval of
the \citet{ChachanEtal2019ajHATP11bPancet} transmission spectra (with
and without the {\Spitzer} observations) we found super-solar metals
mass fractions ($\sim$$10\times$ solar values).  Our results agree
better than \citet{ChachanEtal2019ajHATP11bPancet} with the
extrapolation of the solar-system mass-metallicity
relationship \citep{KreidbergEtal2014apjWASP43bWFC3}, though we remark
that we are estimating metals mass fractions rather than
metallicities.  We suspect that the different conclusions
between our work and those of \citet{ChachanEtal2019ajHATP11bPancet}
arises from the different choice to model the chemistry (free
constant-with-altitude {\it vs.} thermochemical equilibrium).

The case of HAT-P-11b highlights that exoplanet atmospheric retrievals
rely on multiple factors, where the adopted modeling assumptions about
the physics or statistics impact the outcome as significantly as the
observations themselves.  Given the little prior knowledge we have
about any given system, many of these assumptions do not have strong
grounds; they are guesses.  Whether an atmosphere is under
thermochemical equilibrium or not leads to widely different
compositions, and this possibly explains the difference in the
estimated metallicity between our analysis
and \citet{ChachanEtal2019ajHATP11bPancet}.  Whether we trust
stitching of non-simultaneous observations or we apply an uncertainty
correction leads one to favor one (sub)set of observations over
another, reaching different conclusions.  Thus, contrasting multiple
analyses of a given observation can help us to attain a better
understanding not only about planetary physics but also about the
atmospheric retrieval approach.  The tools we have presented here
prepare us to better study the diversity of exoplanet atmospheres that
will be observed in the future.  Hopefully, laboratory advancements
and the development of next-generation telescopes, like the {\JWST}
will enable the study of exoplanets with unprecedented detail.

A compendium of the {\BART} analyses published in this
work is available in a Zenodo repository located at\\
\href{https://doi.org/10.5281/zenodo.5602872}
{https://doi.org/10.5281/zenodo.5602872}.

\acknowledgments

We thank C. Morley and J. Fortney for useful conversations and for
providing radiative-transfer spectra for comparison.  We thank the
anonymous referees for their time and valuable comments.  We thank
contributors to the Python Programming Language and the free and
open-source community (see Software Section below).  We drafted this
article using the aastex6.2 latex
template \citep{AASteamHendrickson2018aastex62}, with some further
style modifications that are available
at \href{https://github.com/pcubillos/ApJtemplate}
{https://github.com/pcubillos/ApJtemplate}.  P.C. was supported by the
Fulbright Program for Foreign Students.  J.B. was supported by the
NASA Earth and Space Science Fellowship, grant NNX12AL83H and the NASA
Exoplanets Research Program, grant NNX17AC03G.  This work was
supported by the Science Mission Directorate's Planetary Atmospheres
Program, grant NNX12AI69G, the NASA Astrophysics Data Analysis Program
grant NNX13AF38G, and the NASA Exoplanets Research Program, grant
NNX17AB62G.  P.R. acknowledges support from CONICYT project Basal
AFB-170002.  Part of this work is based on observations made with the
{\SST}, which is operated by the Jet Propulsion Laboratory, California
Institute of Technology under a contract with NASA.
Part of this
work is based on observations made with the NASA/ESA Hubble Space
Telescope, obtained from the data archive at the Space Telescope
Science Institute.  This research has
made use of NASA's Astrophysics Data System Bibliographic Services.

\software{
{\BART} (\href{https://github.com/exosports/BART}
              {https://github.com/exosports/BART}),
{\mcc} \citep[][]{CubillosEtal2017apjRednoise},
{\TEA} \citep{BlecicEtal2016apsjTEA},
{\repack} \citep{Cubillos2017apjRepack},
\textsc{NumPy} \citep{HarrisEtal2020natNumpy},
\textsc{SciPy} \citep{VirtanenEtal2020natmeScipy},
\textsc{sympy} \citep{MeurerEtal2017pjcsSYMPY},
%\textsc{AstroPy} \citep{Astropy2013aaAstroPy},
\textsc{Matplotlib} \citep{Hunter2007ieeeMatplotlib},
\textsc{IPython} \citep{PerezGranger2007cseIPython},
AASTeX6.2 \citep{AASteamHendrickson2018aastex62},
ApJtemplate (\href{https://github.com/pcubillos/ApJtemplate}
                  {https://github.com/pcubillos/ApJtemplate}),
and
\textsc{bibmanager}\footnote{
\href{https://bibmanager.readthedocs.io}
     {https://bibmanager.readthedocs.io}}
\citep{Cubillos2019zndoBibmanager}.
}

\bibliography{BART2}

\appendix
\section{Analysis of {\Spitzer} Data}
\label{sec:poet}

The {\SST} (warm-mission) obtained four transit light curves of
HAT-P-11b (PI Deming, program ID 80128) using the IRAC instrument: two
visits at 3.6 {\microns} (2011 Jul 07 and Aug 15) and two visits at
4.5 {\microns} (2011 Aug 05 and Aug 29).  The telescope observed in
sub-array mode with a cadence of 0.4 s.  We re-analyzed these light
curves with our Photometry for Orbits, Eclipses, and Transits (POET)
pipeline
\citep{StevensonEtal2010natGJ436b, StevensonEtal2012apjSpitzerHD149026b,
  CampoEtal2011apjWASP12b,
  NymeyerEtal2011apjWASP18b, CubillosEtal2013apjWASP8b,
  CubillosEtal2014apjTrES1}.

The POET analysis started by reading the {\Spitzer} basic calibrated
data (BCD) frames ({\Spitzer} pipeline version 18.18.0).  POET
discarded bad pixels, determined the target position on the detector
(fitting a circular, two-dimensional Gaussian function), and
calculated aperture photometry to produce a raw light curve.  For each
event, we tested circular apertures with radii ranging from 1.75 to
4.0 pixels (in 0.25 pixel increments).

POET simultaneously modeled the out-of-transit system flux, eclipse
curve, and telescope systematics with a Levenberg-Marquardt minimizer
and a Markov-chain Monte Carlo routine.  The {\Spitzer} IRAC
systematics include temporal and intra-pixel sensitivity
variations \citep{CharbonneauEtal2005apjTrES1}.  We modeled the
temporal systematic with a set of time-dependent ``ramp''
models \citep[polynomial, exponential, and logarithmic functions, and
combinations of them;][]{CubillosEtal2013apjWASP8b,
CubillosEtal2014apjTrES1}.  We modeled the intra-pixel systematics
with the Bi-Linearly Interpolated Sub-pixel Sensitivity (BLISS) map
model \citep{StevensonEtal2012apjSpitzerHD149026b}.  We used
a \citet{MandelAgol2002apjLightcurves} model for the transit light
curve.  The transit model fit the planet-to-star radius ratio
($R\sb{\rm p}/R\sb{\rm s}$), transit midpoint time, cosine of
inclination, and semi-major axis-to-stellar radius ratio ($a/R\sb{\rm
s}$).  We adopted the same limb-darkening parameters
as \citet{FraineEtal2014natHATP11bH2O}.  We set the BLISS map
grid size to the RMS of the frame-to-frame pointing jitter; changing
the grid size did not impact the transit
depth \citep{SchwartzCowan2017paspTestingBLISS}.

We determined the best-fitting aperture by minimizing the standard
deviation of the normalized residuals (SDNR) between the data and the
best-fitting model.  Both 3.6 {\micron} data sets showed a clear SDNR
minimum at an aperture radius of 3.0 pixels.  Similarly, both 4.5
{\micron} data sets showed a clear SDNR minimum at 2.5 pixels.  In all
cases, the transit depth remained consistent (within $1 \sigma$)
across the apertures.

We determined the best-fitting ramp model by minimizing the Bayesian
Information Criterion (BIC).  The value of BIC between two competing
models (${\cal M}\sb{1}$ and ${\cal M}\sb{2}$) indicates the
fractional probability, $p({\cal M}\sb{2}|D)$, of being the correct
model \citep[see][]{CubillosEtal2014apjTrES1}.
Tables \ref{table:ha011bp11}--\ref{table:ha011bp22} show the
best-fitting ramps for each data set.

{\renewcommand{\arraystretch}{0.9}
\begin{table}[ht]
\centering
\caption{3.6 {\microns} Visit 1 - Ramp Model Fits}
\label{table:ha011bp11}
\begin{tabular*}{\linewidth} {@{\extracolsep{\fill}} lccc}
\hline
\hline
Ramp & $R\sb{\rm p}/R\sb{\rm s}$\tablenotemark{a} & $\Delta$BIC & $p({\cal M}\sb{2}|D)$ \\
\hline
exponential  & 0.05835(19)  &    0.0 &  $\cdots$ \\
linear       & 0.05767(24)  &    2.3 &  0.24     \\
quadratic    & 0.05839(35)  &    2.9 &  0.19     \\
logarithmic  & 0.05766(22)  &   13.4 & $1.2\tttt{-3}$ \\
\hline
\end{tabular*}
\tablenotetext{}{\footnotesize {\bf Notes.} \sp{a} For this and the
following tables, the value quoted in parentheses indicate the
1$\sigma$ uncertainty corresponding to the least significant digits.}
\end{table}

\begin{table}[ht]
\centering
\caption{3.6 {\microns} Visit 2 - Ramp Model Fits}
\label{table:ha011bp12}
\begin{tabular*}{\linewidth} {@{\extracolsep{\fill}} lccc}
\hline
\hline
Ramp & $R\sb{\rm p}/R\sb{\rm s}$ & $\Delta$BIC & $p({\cal M}\sb{2}|D)$\\
\hline
exponential          & 0.05691(28)  &   0.0 & $\cdots$        \\
exponential + linear & 0.05687(26)  &  10.9 & $4.3\tttt{-3}$  \\
logarithmic          & 0.05713(32)  &  20.2 & $4.1\tttt{-5}$  \\
quadratic            & 0.05729(30)  & 105.8 & $1.1\tttt{-23}$ \\
\hline
\end{tabular*}
\end{table}

\begin{table}[ht]
\centering
\caption{4.5 {\microns} Visit 1 - Ramp Model Fits \label{table:ha011bp21}}
\begin{tabular*}{\linewidth} {@{\extracolsep{\fill}} lccc}
\hline
\hline
Ramp     & $R\sb{\rm p}/R\sb{\rm s}$ & $\Delta$BIC & $p({\cal M}\sb{2}|D)$\\
\hline
no ramp     & 0.05798(33) &   0.0  & $\cdots$ \\
linear      & 0.05813(31) &   3.0  & 0.18 \\         
quadratic   & 0.05807(36) &  13.9  & $9.6\tttt{-4}$ \\         
exponential & 0.05814(38) &  14.0  & $9.1\tttt{-4}$ \\  
\hline
\end{tabular*}
\end{table}

\begin{table}[ht]
\centering
\caption{4.5 {\microns} Visit 2 - Ramp Model Fits}
\label{table:ha011bp22}
\begin{tabular*}{\linewidth} {@{\extracolsep{\fill}} lccc}
\hline
\hline
Ramp     & $R\sb{\rm p}/R\sb{\rm s}$ & $\Delta$BIC & $p({\cal M}\sb{2}|D)$\\
\hline
no ramp     & 0.05814(35) &    0.0  & $\cdots$ \\
linear      & 0.05808(33) &   10.9  & $4.3\tttt{-3}$ \\         
quadratic   & 0.05812(35) &   21.9  & $1.8\tttt{-5}$ \\         
exponential & 0.05814(29) &   22.1  & $1.6\tttt{-5}$ \\         
\hline
\end{tabular*}
\end{table}
}

\begin{figure}[tb]
\centering
\includegraphics[width=\linewidth, clip, trim=40 240 60 240]{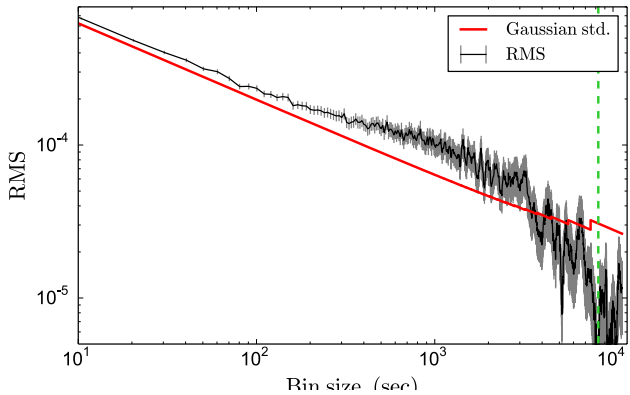}
\caption{Fit residuals' rms (black curve with 1$\sigma$ uncertainties)
  {\vs} bin size for the second visit at 3.6 {\microns}.  The red
  curve shows the expected rms for Gaussian (uncorrelated) noise.  The
  green dashed vertical line marks the transit duration time.}
\label{fig:RMS}
\end{figure}
%% SOURCE: /home/esp01/events/hat-p-11b-patricio/plots/RMS_ha011bp12.py

% Light curves:
\begin{figure*}[tb]
\centering
\includegraphics[width=\textwidth, clip, trim=0 250 0 250]{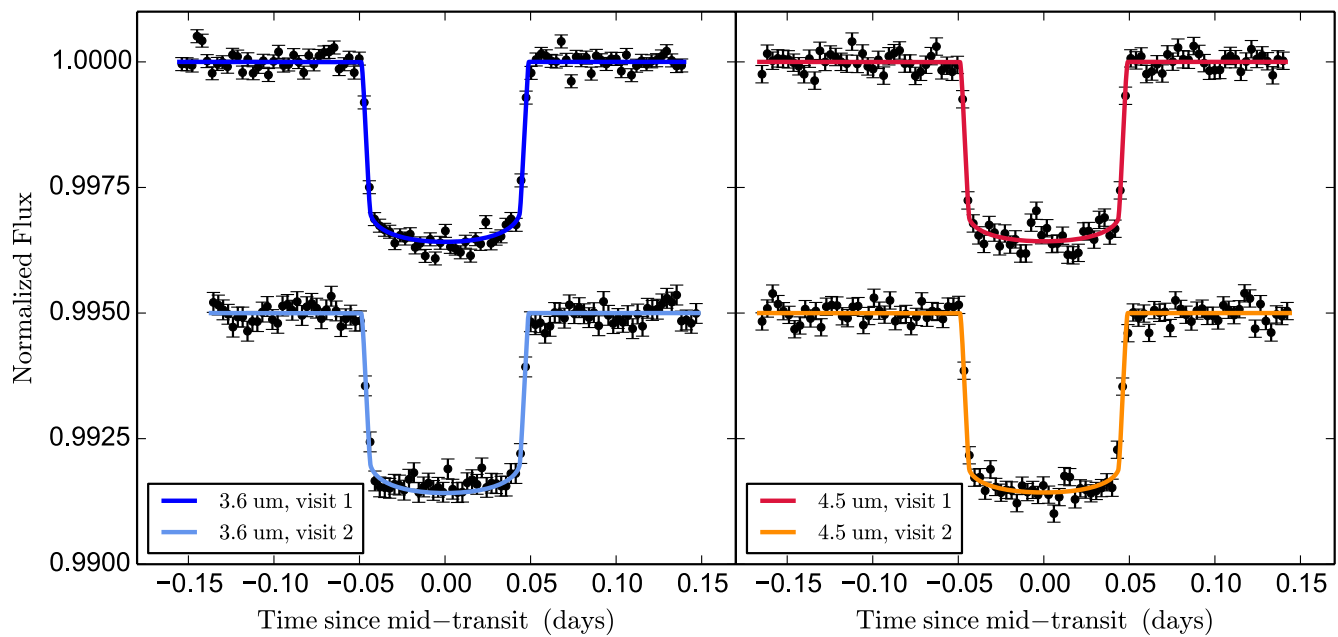}
\caption{
  Normalized, systematics-corrected {\Spitzer} HAT-P-11b transit light
  curves (black points) with the best-fitting models (colored solid
  curves).  The error bars denote the 1$\sigma$ uncertainties.  For
  clarity, we binned the data points and vertically shifted the
  curves.}
\label{fig:c5lightcurves}
\end{figure*}
%% SOURCE: /home/esp01/events/hat-p-11b-patricio/plots/lightcurves.py

At 3.6 {\microns}, the rising-exponential ramp outperformed the other
models in both visits.  Since the second visit at 3.6 {\microns}
showed evidence of correlated noise (Fig.\ \ref{fig:RMS}), we applied
the time-averaging correction factor, and increased the data
uncertainties by a factor of
1.5 \citep[see][]{CubillosEtal2017apjRednoise}.  At 4.5 {\microns},
the no-ramp model outperformed the other models in both visits.

To obtain final transit depths we ran a joint fit combining all four
events.  In this fit, we shared the cosine of inclination and
$a/R\sb{\rm s}$ parameters among all events.  We also shared the
$R\sb{\rm p}/R\sb{\rm s}$ parameter between the events in the same
wavelength band.  Figure \ref{fig:c5lightcurves} shows the
systematics-corrected light-curve data and joint best-fitting model.
Table \ref{table:c5jointfits} summarizes the joint-fit model setup and
results.

The {\Spitzer} transit depth of POET
and \citet{FraineEtal2014natHATP11bH2O} are consistent to each
other within $1 \sigma$.  Furthermore, the depth uncertainties also
agreed, suggesting that both reduction pipelines are statistically
robust.  This is relevant, considering that there have been
disagreements when different groups analyze a same exoplanet light
curve \citep{HansenEtal2014mnrasFeaturelessSpectra}.

{\renewcommand{\arraystretch}{1.0}
\begin{table*}[tb]
\centering
\caption{HAT-P-11b Best Joint-fit Eclipse Light-curve Parameters}
\label{table:c5jointfits}
\begin{tabular*}{\linewidth} {@{\extracolsep{\fill}} lcccc}
\hline
\hline
Parameter                              & 3.6 {\micron} (visit 1)   & 3.6 {\micron} (visit 2)   & 4.5 {\micron} (visit 1)   & 4.5 {\micron} (visit 2)   \\ % OK
\hline
% Centering:
Mean $x$ position (pix)                      &  14.91        &  14.89        &  14.61        &  14.74        \\ % OK
Mean $y$ position (pix)                      &  15.13        &  15.12        &  15.06        &  15.02        \\ % OK
$x$-position consistency\tnm{a} (pix)        & 0.004         & 0.005         & 0.012         & 0.010         \\ % OK
$y$-position consistency\tnm{a} (pix)        & 0.011         & 0.006         & 0.005         & 0.004         \\ % OK
% Photometry:
Aperture photometry radius (pixels)          & 3.0           & 3.0           & 2.5           & 2.5           \\ % OK
% Fitting Results:
System flux \math{F\sb{\rm s}} (\micro Jy)   & 469650(17)    & 468700(17)    & 285075(5)     & 285511(5)     \\ % OK
Transit midpoint (MJD\sb{UTC})\tnm{b}        & 5749.63662(12) & 5788.73918(20) & 5778.96328(15) & 5803.40230(17) \\ % OK
Transit midpoint (MJD\sb{TDB})\tnm{b}        & 5749.63738(12) & 5788.73994(20) & 5778.96404(15) & 5803.40306(17) \\ % OK
Transit duration (\math{t\sb{\rm 4-1}}, hrs) & 2.351(4)      & 2.351(4)      & 2.351(4)      & 2.351(4)      \\ % OK
$R\sb{\rm p}/R\sb{\rm s}$                    & 0.05791(22)   & 0.05791(22)   & 0.05808(25)   & 0.05808(25)   \\ % OK
$\cos(i)$ (deg)                              & 89.52(12)     & 89.52(12)     & 89.52(12)     & 89.52(12)     \\ % OK
$a/R\sb{\rm s}$                              & 16.67(8)      & 16.67(8)      & 16.67(8)      & 16.67(8)      \\ % OK
Limb-darkening coefficient, $c1$             &  0.5750       &  0.5750       &  0.6094       &  0.6094       \\ % OK
Limb-darkening coefficient, $c2$             & $-0.3825$     & $-0.3825$     & $-0.7325$     & $-0.7325$     \\ % OK
Limb-darkening coefficient, $c3$             &  0.3112       &  0.3112       &  0.7237       &  0.7237       \\ % OK
Limb-darkening coefficient, $c4$             & $-0.1099$     & $-0.1099$     & $-0.2666$     & $-0.2666$     \\ % OK
% Nuisance parameters:
Ramp equation ($R(t)$)                       & Rising exponential & Rising exponential   & None          & None          \\ % OK
Ramp, exponential term ($t\sb{0}$)           & $-13.7(5.2)$  & 29.8(3.6)     & $\cdots$      & $\cdots$      \\ % OK
Ramp, exponential term ($r\sb{1}$)           & $-18.3(4.0)$  & 11.5(2.2)     & $\cdots$      & $\cdots$      \\ % OK
% Summary:
Number of free parameters\tnm{c}             & 7             & 7             & 5             & 5             \\ % OK
Total number of frames                       & 62592         & 62592         & 62592         & 62592         \\ % OK
Frames used\tnm{d}                           & 57999         & 56150         & 60962         & 62229         \\ % OK
Rejected frames (\%)                         & 7.34          & 10.29         & 2.60          & 0.58          \\ % OK
BIC                                          & 206185.0      & 206185.0      & 206185.0      & 206185.0      \\ % OK
SDNR                                         & 0.0031105     & 0.0031215     & 0.0042606     & 0.0042751     \\ % OK
Uncertainty scaling factor                   & 0.980         & 1.477         & 1.083         & 1.086         \\ % OK
Photon-limited S/N (\%)                      & 85.3          & 57.0          & 85.9          & 85.6          \\ % OK
\hline
\end{tabular*}
\begin{minipage}{\linewidth}
\footnotesize {\bf Notes.} \sp{a} rms frame-to-frame position difference. \\
\sp{b} MJD = BJD $-$ 2,450,000. \\
\sp{c} In the individual fits. \\
\sp{d} We exclude frames during instrument/telescope
  settling, for insufficient points at a given BLISS bin, and for bad pixels in the photometry aperture.
%% SOURCE: /home/esp01/events/hat-p-11b-patricio/ha011bp11/2015-04-21/run/fgc/ap3000612/joint/2015-05-04_model/2016-03-04_1e5-noramps4.5
\end{minipage}
\end{table*}
}

\end{document}